\documentclass[aoas,preprint]{imsart}

\RequirePackage[OT1]{fontenc}
\RequirePackage{amsthm,amsmath}
\usepackage{amsfonts}
\RequirePackage[authoryear]{natbib}
\RequirePackage[colorlinks,citecolor=blue,urlcolor=blue]{hyperref}
\usepackage{graphicx,xfrac}
\usepackage{bbm}
\usepackage{algorithm}
\usepackage[noend]{algpseudocode}
\usepackage[font={small,it}]{caption}
\usepackage{amsgen,amsmath,amstext,amsbsy,amsopn,amssymb,bbm}
\usepackage{graphicx}
\usepackage{listings}
\usepackage{caption}
\usepackage{subcaption}
\usepackage{natbib}
\usepackage{ulem} 

\newcommand{\D}{\mathbf{D}}
\newcommand{\I}{\mathbf{I}}
\newcommand{\Y}{\mathbf{Y}}
\newcommand{\X}{\mathbf{X}}
\newcommand{\V}{\mathbf{V}}
\newcommand{\W}{\mathbf{W}}

\newcommand{\w}{\mathbf{w}}
\newcommand{\z}{\mathbf{z}}

\newcommand{\Balpha}{\boldsymbol{\alpha}}
\newcommand{\Btheta}{\boldsymbol{\theta}}
\newcommand{\Bbeta}{\boldsymbol{\beta}}

\newcommand{\Btau}{\boldsymbol{\tau}}
\newcommand{\Bgamma}{\boldsymbol{\gamma}}
\newcommand{\BOmega}{\boldsymbol{\Omega}}
\newcommand{\BSigma}{\boldsymbol{\Sigma}}
\newcommand{\1}{\mathbf{1}}
\newcommand{\0}{\mathbf{0}}

\arxiv{1901.08117}

\startlocaldefs
\numberwithin{equation}{section}
\theoremstyle{plain}

\endlocaldefs

\begin{document}

\begin{frontmatter}
\title{Spatial Modeling of Trends in Crime over Time in Philadelphia}
\runtitle{Spatial Modeling of Crime Trends}

\begin{aug}
\author{\fnms{Cecilia} \snm{Balocchi}} \and
\author{\fnms{Shane T.} \snm{Jensen}} 
%\and
%\author{\fnms{Rachel} \snm{Thurston}} 

\affiliation{University of Pennsylvania}

\address{Cecilia Balocchi\\
Department of Statistics \\
The Wharton School \\ 
University of Pennsylvania \\
400 Jon M. Huntsman Hall \\
3730 Walnut Street\\
Philadelphia, PA 19104\\
Email: \href{mailto:balocchi@wharton.upenn.edu}{balocchi@wharton.upenn.edu}}

\address{Shane T. Jensen\\
Department of Statistics \\
The Wharton School \\ 
University of Pennsylvania \\
400 Jon M. Huntsman Hall \\
3730 Walnut Street\\
Philadelphia, PA 19104\\
Email: \href{mailto:stjensen@wharton.upenn.edu}{stjensen@wharton.upenn.edu}}

%\address{Rachel Thurston\\
%Stantec \\ 
%1500 Spring Garden Street\\
%Suite 1100\\
%Philadelphia, PA 19130\\
%Email: \href{mailto:Rachel.Thurston@stantec.com}{Rachel.Thurston@stantec.com}}

\end{aug}

\begin{abstract}
Understanding the relationship between change in crime over time and the geography of urban areas is an important problem for urban planning.  Accurate estimation of changing crime rates throughout a city would aid law enforcement as well as enable studies of the association between crime and the built environment.  Bayesian modeling is a promising direction since areal data require principled sharing of information to address spatial autocorrelation between proximal neighborhoods.  We develop several Bayesian approaches to spatial sharing of information between neighborhoods while modeling trends in crime counts over time.  We apply our methodology to estimate changes in crime throughout Philadelphia over the 2006-15 period, while also incorporating spatially-varying economic and demographic predictors.   We find that the local shrinkage imposed by a conditional autoregressive model has substantial benefits in terms of out-of-sample predictive accuracy of crime.  We also explore the possibility of spatial discontinuities between neighborhoods that could represent natural barriers or aspects of the built environment.  
\end{abstract}


\end{frontmatter}

\section{Introduction}\label{intro}

Modeling and prediction of crime has always been of interest to local authorities, police departments and governments to assure safety of the population and more efficient law enforcement.  Recent availability of detailed crime data has made this effort even more accessible to statistical practitioners and the general public.   

As an example, the Philadelphia police department has released detailed information about reported crimes committed from 2006 to the present day\footnote{http://www.phlcrimemapper.com/}.   The information about each reported crime includes the type of crime (which we will describe in Section~\ref{data}), the date and time of the crime and the GPS location of the crime.   

Using their reported crime data, many police departments have used statistical modeling procedures and algorithms to help predict locations of crimes for better prevention and faster intervention \citep{Hvi16}.   The modeling of crime locations is not only useful for law enforcement but also for marketing strategies related both to real estate and commercial activities, e.g. Trulia\footnote{https://www.trulia.com/} uses crime data as part of their evaluation of the relative safety and attractiveness of different neighborhoods.   In this paper, we will focus on estimating changes in violent crimes over the past decade at a local neighborhood resolution which will involve both temporal and spatial modeling of crime.  

Many different approaches have been taken to the modeling of the spatial distribution of crime.  These approaches can be subdivided into two general categories, either modeling crime as a spatial point process using the specific locations of each reported crime \citep{mohler2011self, taddy2010autoregressive, flaxman2014general} or modeling crime as {\it areal} data, i.e. totals aggregated within larger regions, as in \cite{aldor2016spatio}, \cite{law2014bayesian} and \cite{li2014space}.

A common method for modeling spatial point processes is kriging or Gaussian process interpolation \citep{stein2012interpolation, cressie1990origins}. This can be studied either with a classical approach%, minimizing the mean squared error for prediction, 
, or with a Bayesian approach \citep{banerjee2014hierarchical}.
Alternative popular models consider other frameworks such as Gibbs point processes, Poisson processes and Cox processes; see \cite{moller2007modern}. %\citep{williams1998prediction}. 

Common classical methods for modeling areal data are spatial autoregressive models, that include the Simultaneous Autoregressive Model \citep{whittle1954stationary}, 
%the first-order spatial autoregressive model {\bf needs ref}, 
the spatial Durbin model \citep{anselin1998spatial} and the Conditionally Auto Regressive model \citep{besag1974spatial}; 
%another popular method is the Geographically Weighted Regression \citep{brunsdon1998geographically}. 
for a review of these and other methods see \cite{lesage2009introduction}. 
Many of these models have also been considered and used in a Bayesian framework \citep{banerjee2014hierarchical}.
%Conditionally Auto Regressive model \citep{besag1974spatial} and the Simultaneous Auto Regressive model \citep{whittle1954stationary}; these models are often used in a Bayesian framework \citep{banerjee2014hierarchical}. Other popular methods include the Geographically Weighted Regression \citep{brunsdon1998geographically}.

Our goal in this paper is the estimation of trends in violent crime over the past decade at a high resolution local neighborhood level throughout the city of Philadelphia.   As it is well established that crime frequencies are spatially correlated \citep{herbert1982geography, brantingham1984patterns}, %zhao2017modeling
  we need to create a model that allows the change in crime over time to be correlated by locally proximal neighborhoods.   Our model will also account for characteristics of each local neighborhood, including the population count of the area and economic health of residents, as measured by median income and poverty level of households.   

In addition to aiding law enforcement, accurate estimation of changes in crime at the local neighborhood level would also enable the study of the association between crime trends and changes in the built environment.  We are particularly interested in how aspects of the built environment encourage {\it vibrancy}, a measure of positive human activity, and how vibrancy is associated with safety in local neighborhoods \citep{HumJenSma17}.   

The city of Philadelphia is a particularly interesting case study for estimating trends in crime as it is a large urban area that is currently undergoing substantial development and experiencing population growth for the first time in decades. In addition to our primary goal of estimation of changes in crime in Philadelphia neighborhoods, this application also provides an interesting spatio-temporal data context for comparing different Bayesian shrinkage approaches to spatial areal modeling.    

We will take an areal approach to modeling crime since our primary goal is greater understanding of evolving crime dynamics at the local neighborhood level within the city of Philadelphia.  Our areal units will be U.S. Census block groups which consist of 10-20 city blocks and which are naturally interpretable as “neighborhoods”.   U.S. Census block groups are also the highest resolution for which economic data is available as covariate information.  

Compared to previous areal approaches (e.g. \cite{aldor2016spatio}, \cite{law2014bayesian} and \cite{li2014space}), we are using smaller areal units and we will 
focus on not only total crime but also the trend in crime over time within each local neighborhood.  We have a longer time period (ten years) of recorded crimes for estimating time trends than \cite{law2014bayesian} that worked with property crimes over a two year period.    

Our methodological contribution is the development of a Bayesian spatial modeling framework to explore global vs. local smoothing for our parameter estimates while also allowing for data-driven discontinuities in our model between proximal areal units.  Using a Bayesian approach allows us to induce this smoothing through shrinkage priors for our parameters and also enables us to estimate borders between neighborhoods that have a high probability of being barriers.   

In Section~\ref{data}, we provide details for the neighborhood structure of Philadelphia and describe the detailed crime data that we will use to estimate changes in crime over the past decade.  We also outline the demographic, economic and land use measures we will use as neighborhood-level predictors of violent crime in our spatial models. 
The code for acquiring and cleaning the data that were used in this analysis is available as a GitHub repository at \texttt{https://github.com/cecilia-balocchi/Urban-project}.
In Section~\ref{methods}, we develop several Bayesian modeling approaches for global or local sharing of information between Philadelphia neighborhoods, as well as a model extension that allows for spatial discontinuities in our parameter estimates between proximal neighborhoods.  We then compare these modeling options in terms of both in-sample and out-of-sample predictive accuracy in Section~\ref{sec:comp}.   We visualize and discuss the results of our spatial modeling of crime trends for Philadelphia in Section~\ref{results-application} and then conclude with a brief discussion in Section~\ref{discussion}.  

\section{Population, Economic and Crime Data in Philadelphia} \label{data}

The population and economic data are provided by the US Census Bureau whereas crime data is provided by the Philadelphia Police Department.  Our definition of local neighborhoods in Philadelphia will be based upon the ``block group" geographical units defined by the US Census Bureau.   The city of Philadelphia is divided into 384 census tracts which are divided into 1336 block groups.  
Shapefiles from the US Census Bureau give the boundaries and area of each census block group.  Figure S1 in our supplementary materials \citep{supplement} gives a map outlining the 1336 block groups in Philadelphia.    

Our motivation for analyzing trends in crime at this resolution is two-fold: a. US census block groups consist of 10-20 city blocks which generally matches our concept of a ``neighborhood" and b. the block group level is the highest resolution of the economic data that we will use as predictors of crime.   The average size of block groups in Philadelphia is 0.26 km$^2$, with an average population of 1142 residents.

Our population data was pulled from the census website\footnote{https://factfinder.census.gov/} by setting the geography as all blocks in Philadelphia and setting the data source as ``Hispanic or Latino Origin By Race'' (which is SF1 P5 in their database). The raw demographic data gives the population count in each block group from the 2010 census. Figure S1 in our supplementary materials \citep{supplement} gives the population count for each block group in Philadelphia.  

The same data also has the population count in each block group divided by ethnic categories\footnote{The ethnic categories are: White, Black, Asian, Native Americans, Native Pacific Islanders (including Hawaii), Other, Two or more races (nonhispanic) and Hispanic/Latino.  We combined Native Americans, Native Pacific Islanders, and Two or more races into the Other category, which leads to five ethnicities in our analysis: 1. White, 2. Black, 3. Hispanic, 4. Asian, and 5. Other.}.   From these ethnicity counts, we calculate a measure of the {\it segregation} in each block group as
\[
\text{segregation}_i = \tfrac{1}{2} \sum_{r} | p_{i,r} - \overline{p}_r | 
\]
where $p_{i,r}$ is the proportion of ethnicity $r$ in block group $i$ and $p_{r}$ is the proportion of ethnicity $r$ across the entire city of Philadelphia.  The fraction $\tfrac{1}{2}$ scales this segregation measure to be between 0 and 1.  

In addition to population count and our segregation measure, we will also consider several measures of the economic health of each neighborhood.  Our economic data comes from the American Community Survey from the same US census website as our population data, specifically tables B19301 for income and C17002 for poverty, both from 2013.  This data is only available at the resolution of census block groups.  For each block group (neighborhood) in Philadelphia, we have income per capita as one predictor of crime.  

We also have information about the proportion of households in various states of poverty.   Specifically, we have the fraction of the population in seven different brackets of income-to-poverty-line ratios: $[0, 0.5)$, $[0.5, 1)$, $[1, 1.25)$, $[1.25, 1.5)$, $[1.5, 1.85)$, $[1.85, 2)$, $[2, \infty)$.  For example, the $[0.5, 1)$ bracket represents families that have income between $50\%$ of the poverty line and the poverty line itself.   The poverty line is defined by the Census Bureau according to the size and composition
of a household (e.g. a family with two children has a poverty line threshold of \$23,999).

We use this poverty data to create a single measure of poverty for each block group (neighborhood) by calculating a weighted sum of the proportion of households in each of the seven poverty brackets:
\[
\text{poverty}_i = \sum_{j = 1}^7 w_j \, q_{i,j}
\]
where $q_{i,1}$ is the proportion of households in block group $i$ that are in the lowest bracket $[0, 0.5)$ and $q_{i,7}$ is the proportion of households in block group $i$ in the highest bracket $[2, \infty)$.   We use linearly decreasing weights $\w = [1, 5/6, 4/6, 3/6, 2/6, 1/6, 0]$ to give higher weight to the brackets with higher poverty.   Our poverty measure varies from 0 to 1, with larger values implying higher poverty.

In addition to the demographic and economic predictors described above, we also derive measures of the {\it built environment} that may also be predictive of crime.  Our data on the built environment comes from the zoning designation of each lot in Philadelphia.  Zoning data from the City of Philadelphia provides the area and registered land use designation (e.g. commercial, residential, industrial, vacant, transportation, park, civic) of all 560,000 lots in Philadelphia.  

We create several land use metrics from these zoning designations that could be predictive of crime. First, we calculate the fraction of area in each block group $i$ that is designated as `Vacant',   
\[
\text{vacancy}_{i} = \frac{Area_i (\text{Vacant})}{Area_i}
\]
Second, we calculate the ratio of the area in each block group $i$ that is commercial versus residential, 
\small
\[
\text{comresprop}_i = \frac{Area_i(\text{Commercial})}{Area_i(\text{Commercial}) + Area_i(\text{Residential})}
\]
\normalsize

To summarize, we have created six neighborhood characteristics that we will use as predictors of crime: population count, segregation, median household income, poverty, vacant proportion and commercial vs. residential proportion.   
Some block groups in Philadelphia have missing values for the economic predictors due to a very small or zero population count. We exclude these block groups (a total of eight) from our analysis. We additionally exclude one block group containing the detention centers in Philadelphia. %Block group 1, Census Tract 9891 (in the order in the data is blockgroup 1336)

Our crime data comes from the Philadelphia Police Department and includes all crimes reported by the police in the city of Philadelphia from January 1, 2006 to December 31, 2015.   For each reported crime, we have the type of crime, the date and time of the crime, and the location of the crime in terms of the GPS latitude and longitude (WGS84 decimal degrees).   Each crime in our dataset is categorized into one of several types: homicide, sex crime, armed robbery, assault, burglary, theft, motor vehicle theft, etc. 

We make a distinction between {\it violent} and {\it non-violent} (property) crimes in our analysis.  As defined by the Uniform Crime Reporting program of the \cite{UCR}, {\it violent} crimes include homicides, rapes, robberies and aggravated assaults whereas {\it non-violent} crimes include burglaries, thefts and motor vehicle thefts.

Our own crime categorization differs from the FBI in two ways.   We combine `rapes' and `sex assaults' (which changed in definition in 2013) into a broader `sex crimes' category and consider all `sex crimes' as violent crimes. The FBI also makes a distinction between `aggravated assaults' and `other assaults', with the latter being where an injury does not occur but the threat of injury is present.  In contrast, we combine both `aggravated assaults' and `other assaults' into a broader `assaults' category and consider all `assaults' as violent crimes.  

For this paper, we focus entirely on the modeling of violent crimes as they have the most direct impact on human safety and the perception of safety.   However, non-violent crimes are also important to track for law enforcement and are a focus of ongoing research.  In the subsequent analyses in this paper, we will use `crime' to mean only violent crimes.  

\begin{figure}
\centering
\includegraphics[scale=0.075]{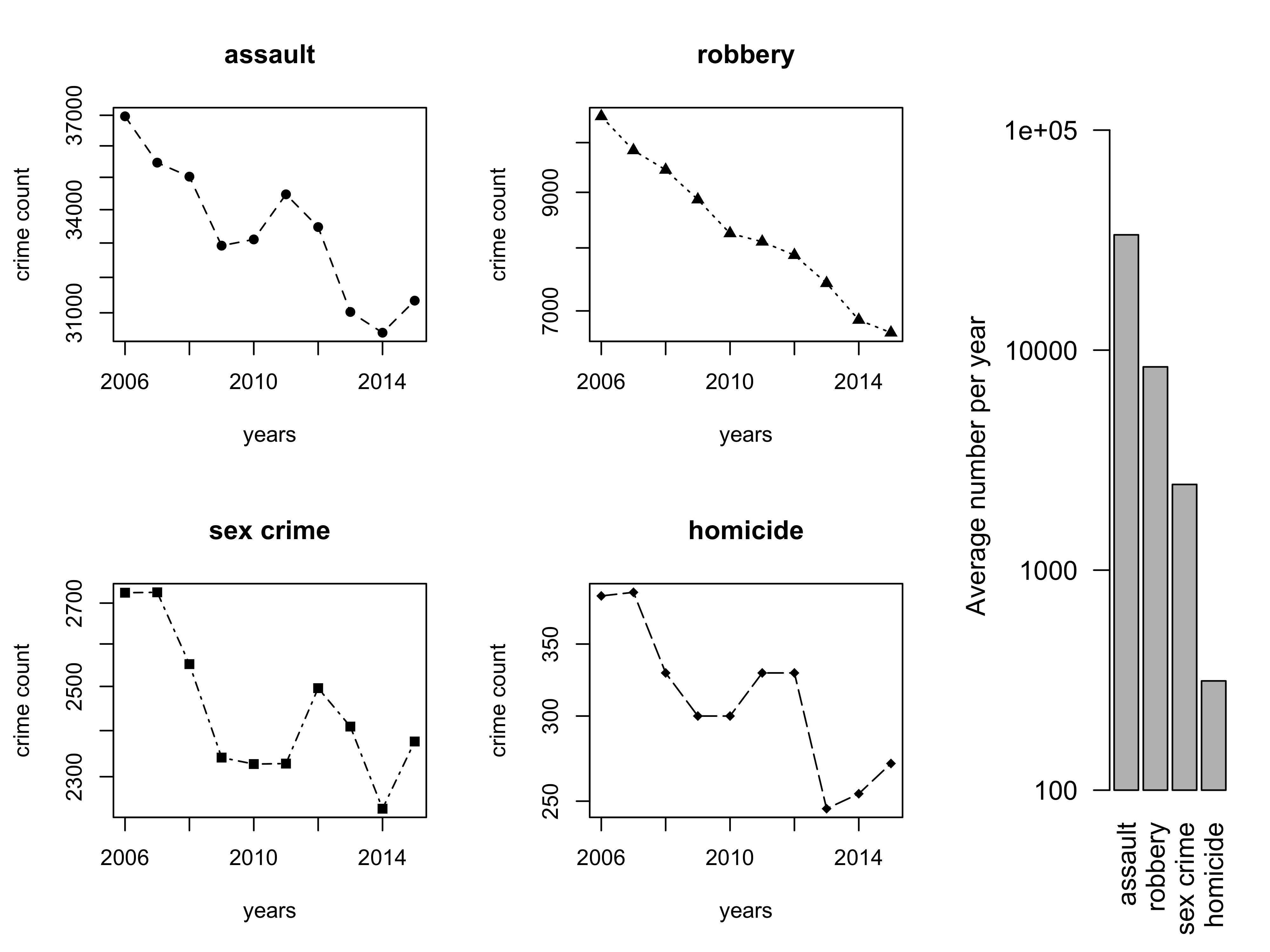} %bar_plot_new2
\caption{Counts of the different types of violent crimes in each year aggregated over the entire city of Philadelphia.} \label{crimebytype}
\end{figure}

In Figure~\ref{crimebytype}, we give the counts of each type of violent crime within each year in 2006-2015, aggregated over the entire city.  We see generally decreasing trends within the assault and robbery categories, which are the most numerous types of crimes.  Sex crimes and homicides are also somewhat decreasing over this time span though it is harder to see this trend given the low counts for either type of crime.   

Clearly, the impression given from Figure~\ref{crimebytype} is that violent crimes are generally decreasing in the city of Philadelphia over the time period from 2006 to 2015.   However, are there specific neighborhoods that show substantially larger decreases or even some 
neighborhoods that show increases in violent crimes in this period?

As discussed in Section~\ref{intro}, we will model the spatial distribution of crime with an areal approach where our areal units are U.S. Census block groups which we define as the local neighborhoods of Philadelphia.  Violent crimes are aggregated within each U.S. Census block group based on the GPS coordinates of each reported crime.    

One issue with this approach is that some crimes occurring near to a boundary between U.S. Census block groups could be aggregated into the incorrect areal unit due to measurement error or ambiguity in their recorded point locations.  This possibility is one of several motivations for our hierarchical Bayesian modeling approach that shares information between adjacent block groups when estimating crime totals and trends in crime over time across the city of Philadelphia.

In Figure S2 of our supplementary materials \citep{supplement}, we give the count of violent crimes per year in each block group averaged over the years 2006-2015.  One can see substantial heterogeneity across block groups in the average counts of violent crimes per year.   There are several outlying values: particular block groups that have much higher average violent crime counts.  

These outlying neighborhoods motivate us to examine violent crime totals on the log scale.  In Figure S2 of our supplementary materials \citep{supplement}, we also give the average of the logarithm of the count of violent crimes per year in each block group, averaged over the years 2006-2015.  We can see more details of the spatial distribution of violent crime on the log scale.  Modeling crime on the log scale has the additional benefit that changes in log crime can be interpreted as percentage changes in crime. 

We also see in Figure S2 evidence of spatial correlation in violent crime totals between proximal block groups throughout the city.  This is not surprising since the factors that lead to crime likely vary throughout the city in a (mostly) spatially continuous fashion.   It is this spatial correlation that will be the focus of our modeling work in Section~\ref{methods}.  

To get an idea of the strength of this spatial correlation, one of the standard statistics used for areal data is Moran's \textit{I} \citep{moran1950notes, banerjee2014hierarchical}, which is defined as 
$$
I = \frac{n}{\sum_i \sum_j w_{ij}} \frac{\sum_i \sum_j w_{ij} (X_i - \bar{X})(X_j - \bar{X})}{\sum_i (X_i - \bar{X})^2}
$$
where $W = (w_{ij})$ is a matrix of weights that capture the spatial proximity of the areal regions. We set $w_{ij}$ to be 1 if block groups $i$ and $j$ share a border and $0$ otherwise.  We use the \textit{queen} contiguity method so two block groups share a border if they share at least a point on their boundaries.  

Moran's \textit{I} can be used for testing for spatial autocorrelation: under the null hypothesis of no spatial association, we can compute exactly the mean (equal to $-\frac{1}{n-1}$) and standard error of Moran's \textit{I}.  Calculating \textit{I} on the total number of violent crimes from 2006 to 2015 in our data gives an observed value of 0.335, compared to a null mean of 0.0007 and standard error of 0.0127, which suggests a highly significant amount of spatial autocorrelation in violent crime totals.   

In the next section, we develop several different Bayesian strategies for modeling violent crime over time and spatially between the areal neighborhoods of Philadelphia.   We will fit our models on the violent crime data from 2006 to 2014, leaving data from 2015 for model comparison and evaluation.

\section{Modeling Areal Crime Data over Space and Time} \label{methods}

As described in Section~\ref{data}, the areal units of our analysis are the 1336 US census block groups of Philadelphia (shown in Figure S1 in our supplementary materials \citep{supplement}).  

For the remainder of this paper, we will use the terms ``block group" and ``neighborhood" interchangeably.   The input data for our analysis is the number of violent crimes, $c_{it}$, reported in year $t$ within neighborhood $i$.  Our temporal range is $t=1,\ldots,T$ with $T =10$, for the years 2006-2015 and our spatial range is $i = 1,\ldots,n$ with $n=1336$, for all the block groups in Philadelphia.  

As seen in the violent crime totals (averaged over time) in Figure S2 of our supplementary materials \citep{supplement}, there are some substantial outlying neighborhoods with high violent crime totals relative to most of the city.   These outliers (and general skewness in violent crime totals) motivates us to model violent crime totals on the logarithmic scale.  This strategy has the additional benefit that linear changes over time in the logarithm of violent crime totals can be interpreted as percentage changes in raw violent crime totals. 

However, because there are a small number of neighborhoods with zero crimes in some years, we need to consider a transformation that is defined at zero.  Accordingly, we use the inverse hyperbolic sine transformation \citep{burbudge1988alternative} that is centered to give values approximately equal to the logarithmic transformation.   Specifically, we calculate our transformed violent crime totals as 
\begin{eqnarray}
y_{it} = \log(c_{it} + \sqrt{c_{it}^2+1}) - \log(2)  
\end{eqnarray}
where $c_{it}$ is the total number of violent crimes reported in year $t$ within neighborhood $i$.  A more common solution would be to add a small non-zero value to the counts, e.g. $\log(c_{it}+1)$.  We prefer the inverse hyperbolic sine transformation as it is numerically equivalent to the $\log$ transformation for large counts but is a better approximation than the $\log(c_{it}+1)$ transformation for small counts.

An alternative modeling strategy for count data does not apply a transformation but assumes a Poisson distribution for the counts \citep{law2014bayesian, li2014space, AndRya17}.   The Poisson model would not work since our data is over-dispersed, and the more flexible negative-binomial distribution does not model mean and variance as intuitively as a normal model.  In addition, the normal model is conjugate for the prior distributions we will be considering which eases posterior estimation.

\subsection{Accounting for Neighborhood Level Covariates}

We use a standard linear regression approach to account for the neighborhood-level economic, demographic and land use predictors of crime.  
Our transformed violent crime totals $y_{it}$ are modeled as, 
\begin{equation}
\label{eq:reg_cov}
y_{it} = \alpha +  \z_i^T\Bgamma + e_{it},
\end{equation}
where $\z_{i}$ is the vector of predictor variables for neighborhood $i$ and $\Bgamma$ is the vector of coefficients for those predictor variables, so $ \z_i^T\Bgamma = \sum_{d=1}^6 \gamma_d z_{id}$.  

As outlined in Section~\ref{data}, we have $d = 6$ predictor variables of crime for each neighborhood: population count, segregation, median household income, poverty, vacant proportion and commercial vs. residential proportion.   We used square root transformations of vacant proportion, commercial vs. residential proportion and poverty and a logarithmic transformation of income to give a more linear relationship with the outcome variable.  

Although yearly demographic and economic data is available after 2013, we avoid extrapolating values of the predictors to earlier years by modeling each predictor variable as {\it static} over the ten year period spanned by our crime data.  We examine the estimated partial effects $\Bgamma$ of these economic, demographic and land use predictors in Section~\ref{results-partial-effects}. 

Although there is interest in the partial effects of our crime predictors, our primary interest lies in the temporal trends captured by $e_{it}$ and the spatial correlation in these trends.   
%in trends over time in the residuals $e_{it}$ within each block group $i$ and the spatial correlation in these trends.  
With these time trends, we will be able to answer questions such as `what areas of the city are increasing or decreasing most quickly in terms of safety?'.  

\subsection{Time Trends with No Spatial Correlation}\label{globalshrinkage}

We can add a global linear trend over time into our model, 
\begin{equation}
\label{eq:reg_cov_ab}
y_{it} = \alpha + \z_i^T\Bgamma + \beta \cdot t + \epsilon_{it} \quad {\rm where} \quad \epsilon_{it} \sim {\rm N} (0, \sigma^2)
\end{equation}
where the scalar coefficient $\beta$ can be interpreted as the global percentage change in violent crime over time across the entire city of Philadelphia and $t$ takes on integer values from 1 to 10 to represent the years 2006-2015. 

However, this model with only a global $\alpha$ and $\beta$ does not allow for heterogeneity between different neighborhoods in the overall level of violent crime or trend in violent crime over time.   We can account for this heterogeneity through neighborhood-specific intercepts $\alpha_i$ and slopes $\beta_i$, which give us the model 
\begin{equation}
\label{eq:reg_cov_akbk}
y_{it} = \alpha_i +  \z_i^T\gamma +  \beta_i \cdot t+ \epsilon_{it} \quad {\rm where} \quad \epsilon_{it} \sim {\rm N} (0, \sigma^2).
\end{equation}

However, model (\ref{eq:reg_cov_akbk}) is over-parameterized: in fact, the effect of our static covariates is completely explained by the neighborhood-specific intercepts $\alpha_i$, so the same fit can be achieved by removing the covariates, 
\begin{equation}
\label{eq:reg_y_akbk}
y_{it} = \alpha_i +  \beta_i \cdot t + \epsilon_{it} \quad {\rm where} \quad \epsilon_{it} \sim {\rm N} (0, \sigma^2)
\end{equation}
Nonetheless, we can still estimate the partial effects of the covariates with an equivalent two-stage approach where we first fit $y_{it} = \alpha + \z_i^T\gamma+ e_{it}$ and then fit the estimated residuals with the neighborhood-specific coefficient model, $\hat{e}_{it} = \alpha_i + \beta_i t + \epsilon_{it}$.

%Thus, an equivalent\footnote{We can still estimate the partial effects on our covariates with an equivalent two-stage approach: fit $y_{it} = \alpha + \z_i^T\gamma+ e_{it}$ and then fit the estimated residuals with the neighborhood specific coefficient model, $\hat{e}_{it} = \alpha_i + \beta_i t + \epsilon_{it}$.} model to (\ref{eq:reg_cov_akbk}) is
%\begin{equation}
%\label{eq:reg_y_akbk}
%y_{it} = \alpha_i +  \beta_i t+ \epsilon_{it} \quad {\rm where} \quad \epsilon_{it} \sim {\rm N} (0, \sigma^2)
%\end{equation}

These neighborhood-specific model coefficients allow us to identify regions of Philadelphia with different levels of crime as well as different trends in crime over the past decade.   This richer model is also motivated by fit to the data: a regression model with neighborhood-specific coefficients explains significantly more variation according to an F-test. 

That said, we do not expect that every single neighborhood in Philadelphia would have unique coefficients, so we still risk over-parametrization with this model.   We address this over-parameterization by imposing shared prior distributions for the neighborhood-specific coefficients from our time trend model (\ref{eq:reg_y_akbk}), 
\begin{eqnarray}
\Balpha \,\, {\sim} \,\, {\rm N} \, (\alpha_0 \cdot \1 \, , \, \tau_\alpha^2 \cdot \I) \label{eq:global_shrinkage1} \\
\Bbeta \,\, {\sim} \,\, {\rm N} \, (\beta_0 \cdot \1 \, , \,  \tau_\beta^2 \cdot \I) \label{eq:global_shrinkage2}\\
\Bgamma \,\, {\sim} \,\, {\rm N} \, (\0 \, , \,  \tau_\gamma^2 \cdot \I) \label{eq:global_shrinkage3}
\end{eqnarray}
where we denote our collection of neighborhood specific coefficients with $\Balpha = (\alpha_1, \ldots, \alpha_n)$ and $\Bbeta = (\beta_1, \ldots, \beta_n)$.  $\Bgamma = (\gamma_1, \ldots, \gamma_d)$ collects the coefficients (partial effects) of the predictor variables which are shared by all neighborhoods.   

We complete this model formulation by placing flat priors on the global means $\alpha_0$ and $\beta_0$, $p(\alpha_0,\beta_0) \propto 1$, and inverse gamma priors on the variance parameters 
\begin{align*}
\sigma^2 &\sim \mbox{Inv-Gamma}(a_\sigma, b_\sigma)\\
\tau^2_\alpha &\sim \mbox{Inv-Gamma}(a_\alpha, b_\alpha)\\
\tau^2_\beta &\sim \mbox{Inv-Gamma}(a_\beta, b_\beta)\\
\tau^2_\gamma &\sim \mbox{Inv-Gamma}(a_\gamma, b_\gamma).
\end{align*}
The variance hyper-parameters are tuned in an empirical Bayes fashion so that the prior mean of the variance parameters is equal to the variance estimated from the model with no shrinkage, and the prior variance is small.  Using non-informative priors for these variance parameters produced nearly identical results.  See Section 4 of our Supplementary Materials \citep{supplement} for details.  

This Bayesian hierarchical model shares information between neighborhoods by shrinking the neighborhood specific coefficients $\alpha_i$ and $\beta_i$ towards global parameters $(\alpha_0,\beta_0)$ for the entire city.  For this reason, we refer to this approach as the {\it global shrinkage} model.
 
However, this global shrinkage model does not account for the spatial proximity between neighborhoods when sharing information.  
We expect close neighborhoods to behave similarly while we want distant neighborhoods to be informative but not as directly influential as adjacent ones.  In other words, we may prefer a model that imposes {\it local} shrinkage rather than global shrinkage.  

A model with local sharing of information would also be better able to address the substantial spatial correlation that we see in our application. Testing with Moran's \textit{I} shows that the residuals from the global shrinkage model are significantly spatially correlated.   In the next subsection, we will explore conditional auto-regressive models for local sharing of information. 

\subsection{Time Trends with a Spatial Conditional Auto Regressive Model}\label{spatialCAR}

A popular way of incorporating spatial information is through a prior distribution that is specified according to a Conditional Auto Regressive (CAR) model, which was introduced in its most general formulation by \cite{besag1974spatial}.   The CAR model is a Gaussian Markov random field which induces spatial dependence through an adjacency matrix for the areal units, which in our case are neighborhoods in Philadelphia.  

Several variations of this CAR framework are reviewed and compared in \cite{Lee2011}.  In this paper, we will use the proper CAR formulation introduced by \cite{leroux2000estimation}.

Let $\Btheta$ denote a vector of elements that are potentially spatially correlated, such as our neighborhood-specific intercepts $\Balpha$ or slopes $\Bbeta$.  \cite{leroux2000estimation} defines the distribution of each $\theta_i$ given the other $\theta_{-i}$ as a normal distribution centered at a weighted average of a global mean and the $\theta_j$'s from bordering neighborhoods, 
\begin{equation}
\label{eq:cond_leroux}
\theta_i \mid \Btheta_{-i}, \theta_0, \tau^2 \sim {\rm N} \left( \dfrac{\rho \, \sum_j w_{ij} \, \theta_j + (1- \rho) \, \theta_0}{\rho \, \sum_j w_{ij} + (1-\rho)}, \dfrac{\tau^2}{\rho \, \sum_j w_{ij} + (1-\rho)}\right),
\end{equation}
where $w_{ij}$ are adjacency weights that are equal to 1 if the neighborhoods $i$ and $j$ share a border and equal to 0 otherwise.

We collect these adjacency weights $w_{ij}$ into an adjacency matrix $\W$ that we assume (for now) to be known since we can easily use the shapefiles from the US Census Bureau to determine which of the 1336 neighborhoods (census block groups) share a border. 

For now, we consider these adjacency weights $w_{ij}$ to be fixed.  However, in Section~\ref{varW} we will extend our model to allow those weights to vary since some borders may represent {\it barriers} between neighborhoods (e.g. highways or rivers), in which case we would not want to share information across that particular border.

The parameter $\rho \in [0,1]$ represents the strength of the spatial correlation between the components of $\Btheta$, where larger values of $\rho$ correspond to a stronger influence of bordering neighborhoods.  In the special case of $\rho = 0$, the CAR prior (\ref{eq:cond_leroux}) reduces to the \textit{global shrinkage} prior (\ref{eq:global_shrinkage1})-(\ref{eq:global_shrinkage2}).  

It can be proved \citep[Ch.3]{banerjee2014hierarchical} using Brook's lemma \citep{brook1964distinction}, that the joint distribution of $\Btheta$ is uniquely determined by the set of conditional distributions defined in \ref{eq:cond_leroux}:
\begin{equation}
\Btheta \vert \theta_0, \tau^2 \sim {\rm N} \left( \theta_0  \cdot \1 \, , \, \tau^{2} \cdot [\rho (\D_W-\W) + (1-\rho) \I]^{-1} \right) 
 \label{eq:joint_leroux}
\end{equation}
where $\1$ is a vector of 1's and $\D_W-\W$ is the Laplacian matrix based on our neighborhood adjacency matrix $\W$. For values of $\rho$ in $[0,1)$ the joint distribution is proper, while for $\rho =1$ the distribution is degenerate \citep{Lee2011}.  By adding the constraint $\sum_i(\theta_i - \theta_0)=0$ we can get a distribution for a $n$-dimensional vector, concentrated in a $(n-1)$-dimensional subspace; this is known as the intrinsic CAR by \cite{Besag1991}.

We will employ this CAR model as prior distributions for the vectors of time trend coefficients $\Balpha$ and $\Bbeta$. We assume $\Balpha$ and $\Bbeta$ are {\it a priori} independent.  In vector form, the CAR model (\ref{eq:cond_leroux}) corresponds to the following prior distributions for $\Balpha$ and $\Bbeta$, 
\begin{eqnarray}
\Balpha \,\, {\sim} \,\, {\rm N} \, \left( \alpha_0 \cdot \1 \, , \, \tau_\alpha^2 \cdot \BSigma \right) \label{eq:joint_leroux1} \\
\Bbeta \,\, {\sim} \,\, {\rm N} \, \left(\beta_0 \cdot \1 \, , \,  \tau_\beta^2 \cdot \BSigma \right) \label{eq:joint_leroux2}
\end{eqnarray}
where $\BSigma^{-1} = \rho (\D_W-\W) + (1-\rho) \I$.    

We use the same prior distributions for $\alpha_0$ and $\beta_0$ and our variance parameters as in the {\it global shrinkage} model in the previous subsection.   For the additional spatial parameter $\rho$, we choose a Beta$(10,10)$ prior distribution which has mean equal to $0.5$ and a small variance in order to avoid the endpoints of the interval $[0,1]$.

The posterior distributions for the spatial CAR model and the global shrinkage model (Section \ref{globalshrinkage}) can be implemented via a Gibbs sampler \citep{GemGem84}.  Implementation details are given in our supplementary materials \citep{supplement}.  

\subsection{Allowing Neighborhood Border Weights to Vary} \label{varW}

For most types of areal data, the weights $\W$ that encode the spatial connection between the areal units are considered to be fixed and known.  In our data context, the areal units are neighborhoods and the weights $\W$ encode which neighborhoods share a border and hence induce shrinkage on each other in our spatial CAR models outlined in Section~\ref{spatialCAR}.   

However, within any large city, some borders between neighborhoods consist of natural or artificial {\it barriers} such as rivers, highways or train tracks.  These barriers could reduce the similarity in crime trends between neighborhoods, and so we would not want to shrink estimates across those barriers.  The implication of these barriers for the spatial CAR models in Section~\ref{spatialCAR} are that some weights $w_{ij} = 1$ should really be $w_{ij} = 0$ since those neighborhoods share a border that is actually a barrier.   

Attempting to set which borders should actually be barriers manually would be tedious for a large city and also require extensive domain knowledge and subjective decision making. We instead prefer to infer these barriers from the data by allowing a subset of weights $w_{ij}$ to be random variables in our model.  

Specifically,  we consider the set of indices of pairs of neighborhoods which share a border according to the geography of Philadelphia. The matrix $\W$ is symmetric so the random variables $w_{ij}$ and $w_{ji}$ are considered to be the same object.    We model the $w_{ij}$ for neighborhood pairs that share a border as Bernoulli random variables with an prior probability $\phi$ of $ w_{ij} = 1$.   Any weights $ w_{ij} = 0$ according to the geography of Philadelphia will remain fixed at $ w_{ij} = 0$ since we do not want to form connections between non-proximal neighborhoods. 

We expect {\it a priori} that the probability $\phi$ will be close to 1, since relatively few borders between neighborhoods actually should be barriers. For this reason we choose the prior for $\phi$ to be a ${\rm Beta}(9,1)$ distribution which has mean close to one and small variance.  

Moreover, we expect that the spatial distribution of the neighborhood-specific crime levels $(\alpha_i)$ may be different from the neighborhood-specific trends in crime over time $(\beta_i)$, so we allow for different barriers when we model the distribution of $\Balpha$ and $\Bbeta$.  In particular, we consider two random matrices $\W^\alpha$ and $\W^\beta$ where a subset of the elements of these matrices are random as described above: $w_{ij}^\alpha \, \vert \, \phi^\alpha \sim {\rm Bernoulli}(\phi^\alpha)$ and $w_{ij}^\beta \, \vert \, \phi^\beta \sim {\rm Bernoulli}(\phi^\beta)$ for neighborhood pairs $(i,j)$ that share a border.

These two weight matrices then determine the local shrinkage of our spatial CAR model from the previous subsection:
\begin{eqnarray}
\Balpha \, \vert \, \W^\alpha \,\, {\sim} \,\, {\rm N} \, \left( \alpha_0 \cdot \1 \, , \, \tau_\alpha^2 \cdot \BSigma_\alpha \right) \label{eq:joint_randomW1} \\
\Bbeta \, \vert \, \W^\beta \,\, {\sim} \,\, {\rm N} \, \left(\beta_0 \cdot \1 \, , \,  \tau_\beta^2 \cdot \BSigma_\beta \right) \label{eq:joint_randomW2}
\end{eqnarray}
where $\BSigma_\alpha^{-1} = \rho \cdot (\D_{W^\alpha}-\W^\alpha) + (1-\rho) \I$ and $\BSigma_\beta^{-1} = \rho \cdot (\D_{W^\beta}-\W^\beta) + (1-\rho) \I$. 

Allowing variable border weights can lead to over-parametrization since we are adding as many parameters as the number of borders, which makes the shrinkage imposed by prior parameters $\phi^\alpha$ and $\phi^\beta$ important.  A more sophisticated approach, which is the focus of ongoing work, would be to partition our areal units into clusters with barriers represented as cluster boundaries. 

To implement this extended model with some variable border weights, a step is added to our Gibbs sampler that samples each border weight conditional on the current values of the other model parameters.   Details are given in our supplementary materials \citep{supplement}.  

The idea of detecting discontinuities at boundaries is often referred to as \textit{wombling} after the seminal work of \cite{Womble315} and has been very popular in the disease mapping literature.  However, most papers have approached detection of boundaries as a selection problem that is performed after inference \citep[see, e.g.][]{boots2001using, li2011mining, banerjee2012bayesian, lu2005bayesian, lee2013locally}.   

In contrast, we incorporate the possibility of discontinuities at boundaries directly into our model through variable $\W^\alpha$ and $\W^\beta$, which allows us to incorporate potential barriers into our estimation of neighborhood-specific parameters $\Balpha$ and $\Bbeta$.  \cite{lee2012boundary} and \cite{lu2007bayesian} take a similar approach in the context of disease mapping, but with a more elaborate model for ${\rm P}(w_{ij} = 1)$ that is a function of dissimilarity between covariate values in units $i$ and $j$.

In Section~\ref{sec:comp}, the different models presented in this section are compared in terms of their accuracy of their in-sample and out-of-sample predictive accuracy.  We then visualize the estimated trends in crime over time in Philadelphia and discuss several insights from our results in Section~\ref{results-application}.  

\section{Comparison of Predictive Accuracy} \label{sec:comp}

In the previous section, we outlined a no shrinkage model (Section~\ref{globalshrinkage}) and several hierarchical Bayesian models for estimating the neighborhood-level trend in crime over time, including a global shrinkage model (Section~\ref{globalshrinkage}), a spatial CAR models for local shrinkage (Section~\ref{spatialCAR}), and finally an extension of the spatial CAR model to allow a subset of border weights to vary (Section~\ref{varW}). 

We now compare each of these model alternatives based on several measures of the accuracy of their predictions on both in-sample and out-of-sample hold-out data.  Recall that we have 10 years of crime data for the city of Philadelphia, from the beginning of 2006 to the end of 2015.  We estimate each model using the crime data for the first nine years (2006-2014). 

We assess the {\it in-sample} accuracy of each model by computing the mean squared error of the predictions of violent crime totals for 2014, which is a year that was included in model estimation, 
\begin{eqnarray}
{\rm MSE}_{\rm in} = \frac{1}{1336} \sum\limits_{i=1}^{1336} (y_{i,{\rm 2014}} - \hat{y}_{i,{\rm 2014}})^2.
\end{eqnarray}

We assess the {\it out-of-sample} accuracy of each model by computing the mean squared error of the predictions of violent crime totals for 2015, which is a year that was {\it not} included in model estimation,  
\begin{eqnarray}
{\rm MSE}_{\rm out} = \frac{1}{1336} \sum\limits_{i=1}^{1336} (y_{i,{\rm 2015}} - \hat{y}_{i,{\rm 2015}})^2. \label{outofsampleerror}
\end{eqnarray}

To ensure our evaluation is not overly dependent on any idiosyncratic aspects of the 2015 data, we also calculate the {\it cross-validated out-of-sample} accuracy of each model by calculating the mean square error ${\rm MSE}_{\rm out}^t$ when using year $t$ as the hold out data in the same way that 2015 is used as the hold out data in (\ref{outofsampleerror}), i.e.
\begin{eqnarray}
\phantom{yomom} {\rm MSE}_{\rm cv} = \frac{1}{10} \sum\limits_{t=1}^{10} {\rm MSE}_{\rm out}^t \quad {\rm where} \quad {\rm MSE}_{\rm out}^t = \frac{1}{1336} \sum\limits_{i=1}^{1336} (y_{i,t} - \hat{y}_{i,t})^2. \hspace{-0.5cm} \label{cverror}
\end{eqnarray}

In Table~\ref{tb:cv}, we compare the predictive accuracy of four different models with neighborhood-specific coefficients outlined in Section~\ref{methods}: 1. the time trend model (\ref{eq:reg_y_akbk}) without shrinkage between neighborhoods, 2. the global shrinkage model with priors (\ref{eq:global_shrinkage1}) and (\ref{eq:global_shrinkage2}), 3. the local shrinkage model with spatial CAR priors (\ref{eq:joint_leroux1}) and (\ref{eq:joint_leroux2}) and 4. the local shrinkage spatial CAR model with variable borders (\ref{eq:joint_randomW1}) and (\ref{eq:joint_randomW2}).  For additional reference, we also provide the mean square error for fitting a single trend (``Global $\alpha \, , \, \beta$") across the entire city.

\begin{table}[ht!]
\centering
\begin{tabular}{lcccccc}
Model  & ${\rm MSE}_{\rm in}$ & ${\rm MSE}_{\rm out}$ & $\%$ change & ${\rm MSE}_{\rm cv}$& Moran's $I$\\ 
\hline
Global $\alpha \, , \, \beta$ & 0.3558 & 0.3694 & +182.4 & 0.3043  & - \\
Separate $\alpha_i \, , \, \beta_i$ Models & & & & & \\
\phantom{y} No Shrinkage (\ref{eq:reg_y_akbk})  & 0.0567 &0.1308 & - & 0.1001 & 0.17\\ 
\phantom{y} Global Shrinkage (\ref{eq:global_shrinkage1})-(\ref{eq:global_shrinkage2}) & 0.0698 & 0.1080 & -17.4 & 0.0928 & 0.17 \\ 
\phantom{y} Spatial CAR (\ref{eq:joint_leroux1})-(\ref{eq:joint_leroux2})   &  0.0703 & 0.1052 & - 19.5 & 0.0922  & 0.61\\
\phantom{y} Variable Borders (\ref{eq:joint_randomW1})-(\ref{eq:joint_randomW2})& 0.0706 & 0.1069 & -18.2 & 0.0927 & 0.49 \\
%\phantom{y} Variable Borders - only A & 0.0711 & 0.1050 & - 19.7 & 0.0921 & 0.64 \\
\end{tabular} 
\caption{Comparison of predictive accuracy between the different models outlined in Section~\ref{methods}.  The mean squared error for both in-sample and out-of-sample predictions are provided, as well as the percentage change in ${\rm MSE}_{\rm out}$relative to model (\ref{eq:reg_y_akbk}) without shrinkage.  We also provide the Moran's $I$ measure of spatial correlation calculated on the estimated time trends $\beta_i$ from each model.}
\label{tb:cv}
\end{table}

We see in Table~\ref{tb:cv} that the model with a global trend over time (``Global $\alpha \, , \, \beta$") for the entire city has very poor predictive accuracy compared to the models that allow neighborhood-specific time trends (``Separate $\alpha_i \, , \, \beta_i$").   

Among the neighborhood-specific time trend models, the global shrinkage model has substantially lower out-of-sample mean square errors than the baseline time trend model without any shrinkage between neighborhoods.  The best in-sample mean squared error was achieved by the model without shrinkage, as we expect from the least square method, though at a cost of having the worst out-of-sample accuracy. 

The model with local shrinkage via the spatial CAR prior further reduces the out-of-sample mean square errors compared to the global shrinkage model.   The model that allows variable borders does not further improve the out-of-sample mean squared errors, though we explore in Section~\ref{results-borders} that it helps with the interpretation.

Table~\ref{tb:cv} also provides Moran's $I$ measure of spatial autocorrelation, calculated on the posterior mean of the neighborhood-specific time trends ($\beta_i$'s).   We see that the spatial CAR model induces a larger spatial correlation in the $\beta_i$'s than the models with global shrinkage or without shrinkage.  The local shrinkage model has a Moran's I value of 0.61 ({\it s.e.} = 0.016), which suggests there is substantial spatial autocorrelation in the change in crime within Philadelphia.

In summary, allowing for local shrinkage of the neighborhood-specific crime trend coefficients via the spatial CAR priors (\ref{eq:joint_leroux1}) and (\ref{eq:joint_leroux2}) leads to the best out-of-sample predictive accuracy.  In Section~\ref{results-application}, we visualize the parameters of this model and discuss the implications of these results for crime in Philadelphia.

Although the variable border model extension does not improve out-of-sample predictive accuracy, we will also see in Section~\ref{results-application} that visualizing the borders that have been turned into barriers by this model provide insight into discontinuities in crime trends in the city of Philadelphia.

\section{Interpretation of Model Parameters} \label{results-application}

In Section~\ref{results-partial-effects}, we examine the estimated partial effects for the static predictor variables created from the data outlined in Section~\ref{data}.   We then visualize and compare the estimated neighborhood-specific levels ($\alpha_i$'s) and time trends ($\beta_i$'s) on crime from our different models in Section~\ref{results-crimetrends}.   In Section~\ref{results-borders}, we examine the results from our model extension outlined in Section~\ref{varW} that allows a subset of neighborhood borders in Philadelphia to be estimated as barriers.  Finally, in Section~\ref{results-bestworst} we discuss the neighborhoods with the most extreme levels and changes in crime over time over the past ten years in Philadelphia.

\subsection{Partial Effects of Static Predictors} \label{results-partial-effects}

\begin{figure}
\centering
\includegraphics[scale=0.15]{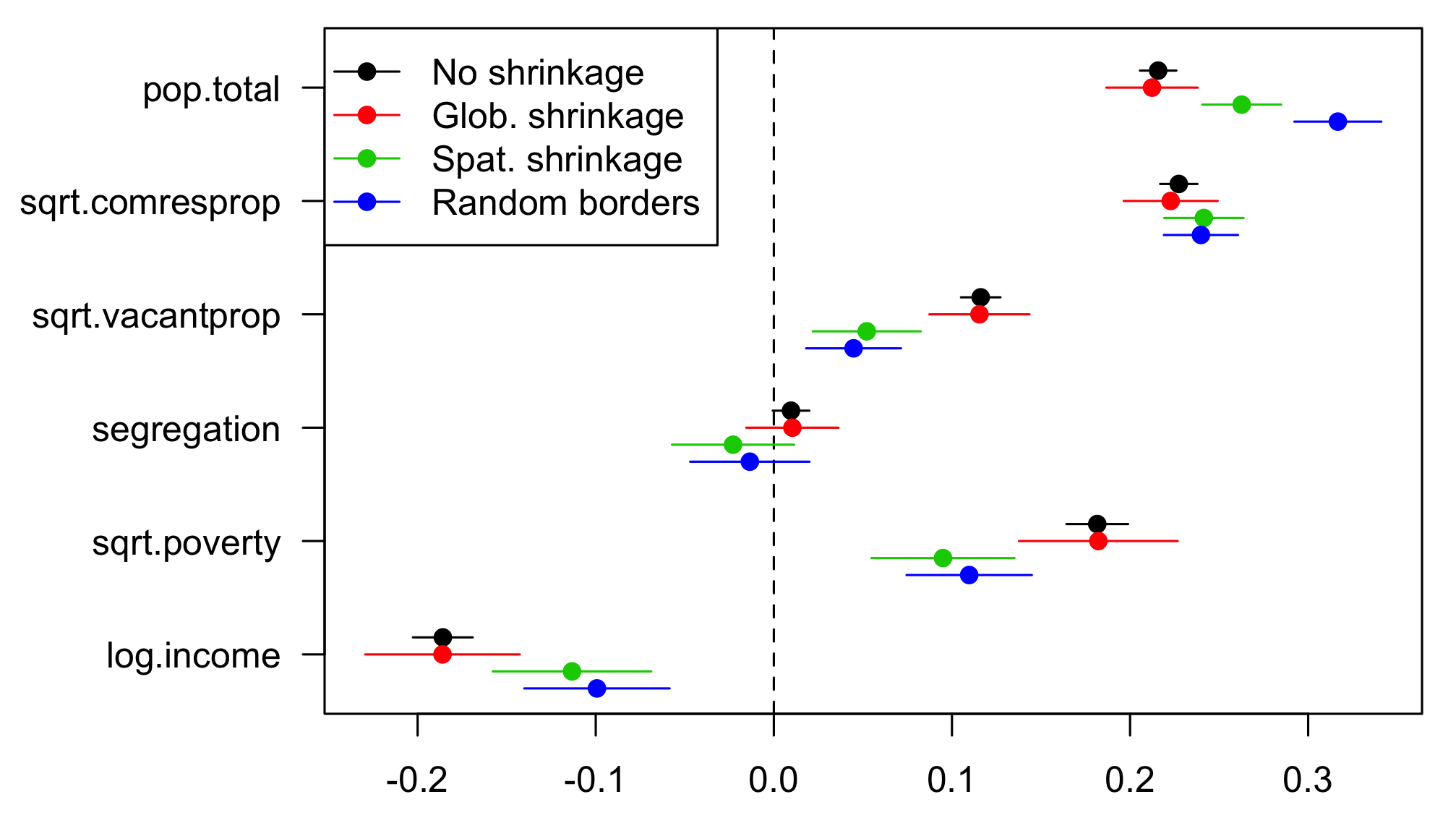} %bar_plot_new2
\caption{Estimated partial effects $\gamma_d$ from four different models indicated in the legend.  For the no shrinkage model, we plot the maximum likelihood estimate and 95\% confidence interval.  For three Bayesian shrinkage models, we plot the posterior mean and 95\% posterior interval.} \label{fig-partial-effects}
\end{figure}

Figure~\ref{fig-partial-effects} gives the estimated partial effects $\gamma_d$ for each static predictor variable $d$ from the four models outlined in Section~\ref{methods}.  We provide additional numerical details in Table S1 of our supplementary materials \citep{supplement}.

We see that among the six predictor variables created in Section~\ref{data}, only the segregation measure is not a significant predictor of crime.  All predictor variables are on the same scale and so we can directly compare the values of their partial effects.

We see that the strongest predictors of crime are total population and the commercial versus residential proportion, with more populated and more commercial neighborhoods being associated with higher crime.  Income and poverty are also significantly predictive of violent crimes but we must be more cautious about interpreting these partial effects given the high collinearity between income and poverty.  Each of these observations on the partial effects $\Bgamma$ is relatively consistent across the four models outlined in Section~\ref{methods}.  

\subsection{Visualizing Neighborhood-Specific Coefficients} \label{results-crimetrends}

Our primary interest in terms of interpretation are the estimated neighborhood-specific coefficients, $\alpha_i$'s and $\beta_i$'s, that represent the level of violent crimes and change in violent crimes over time in Philadelphia, respectively.   

\begin{figure}
\hspace*{-1in}
\includegraphics[scale = 0.14]{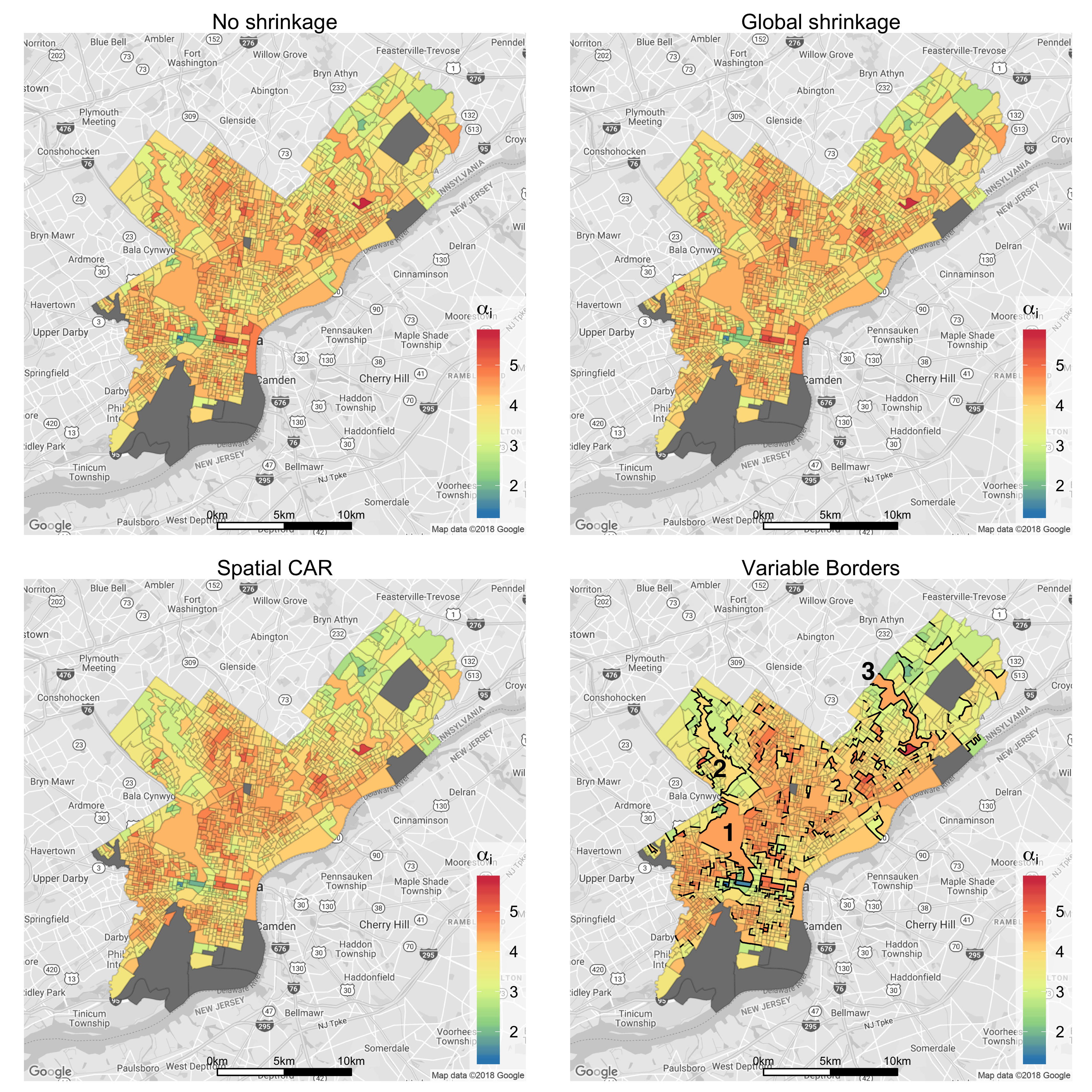}
\caption{Maps of Philadelphia colored by the estimated intercept from our four different models. {\bf Top-left:} Maximum likelihood estimates of $\alpha_i$ from the no shrinkage model (\ref{eq:reg_y_akbk}). {\bf Top-right:} Posterior means of $\alpha_i$ from the global shrinkage model (\ref{eq:global_shrinkage1})-(\ref{eq:global_shrinkage2}). {\bf Bottom-left:} Posterior means of $\alpha_i$ from the spatial CAR model (\ref{eq:joint_leroux1})-(\ref{eq:joint_leroux2}). {\bf Bottom-right:} Posterior means of $\alpha_i$ from the spatial CAR model with variable borders (\ref{eq:joint_randomW1})- (\ref{eq:joint_randomW2}). The black lines represent borders turned into barriers.  These maps were created with the R package {\tt ggmap} \citep{kahle2013ggmap}. }\label{fig-alpha}
\end{figure}

\begin{figure}
\hspace*{-1in}
\includegraphics[scale = 0.14]{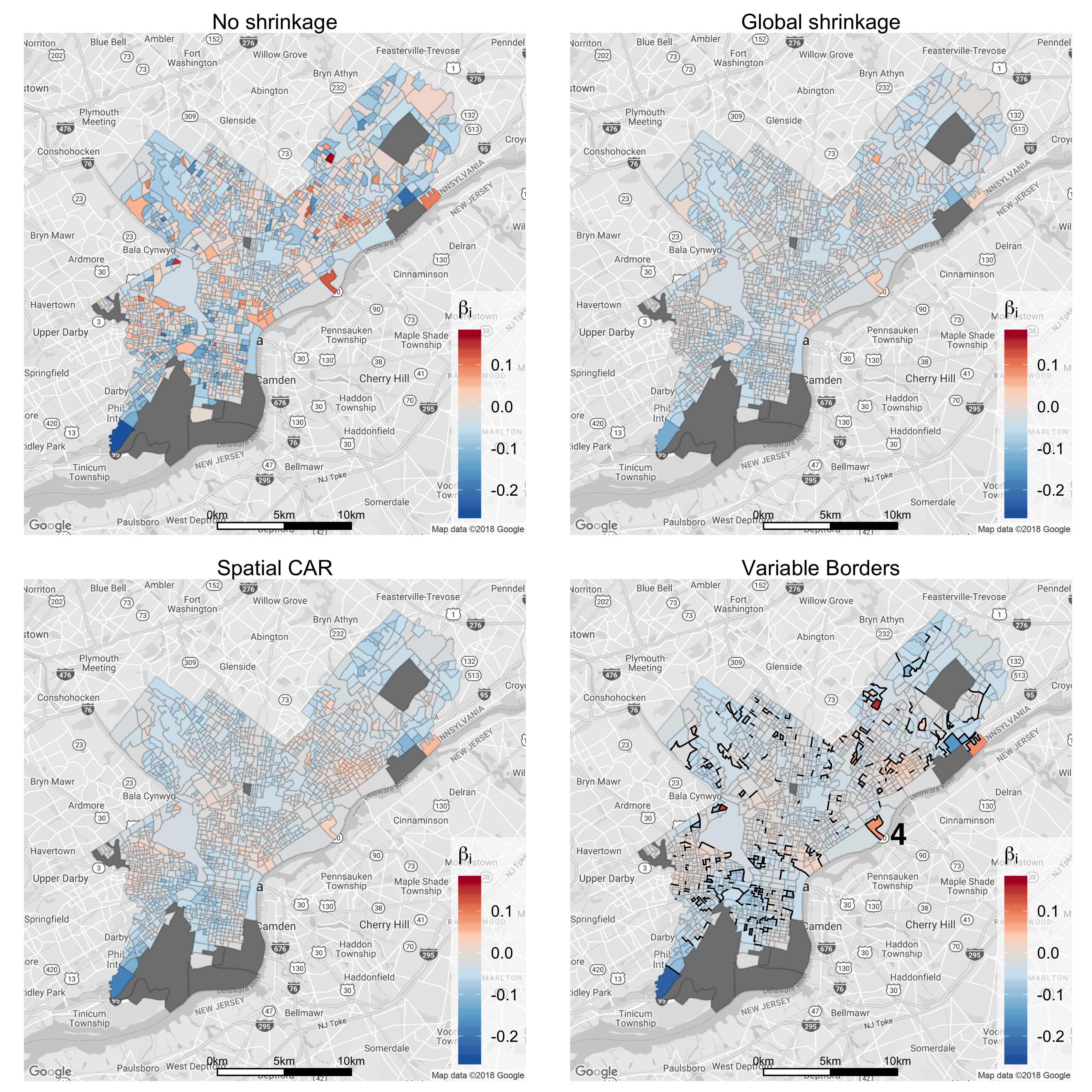}
\caption{Maps of Philadelphia colored by the estimated slope on time from our four different models. {\bf Top-left:} Maximum likelihood estimates of $\beta_i$ from the no shrinkage model (\ref{eq:reg_y_akbk}). {\bf Top-right:} Posterior means of $\beta_i$ from the global shrinkage model (\ref{eq:global_shrinkage1})-(\ref{eq:global_shrinkage2}). {\bf Bottom-left:} Posterior means of $\beta_i$ from the spatial CAR model (\ref{eq:joint_leroux1})-(\ref{eq:joint_leroux2}). {\bf Bottom-right:} Posterior means of $\beta_i$ from the spatial CAR model with variable borders (\ref{eq:joint_randomW1})- (\ref{eq:joint_randomW2}). The black lines represent borders turned into barriers.  These maps were created with the R package {\tt ggmap} \citep{kahle2013ggmap}. } \label{fig-beta}
\end{figure}

In Figures~\ref{fig-alpha} and~\ref{fig-beta}, we give maps where each block group in Philadelphia is colored by the estimated neighborhood-specific levels of crime $\hat{\alpha}_i$ and changes in crime over time $\hat{\beta}_i$ respectively, from the four models outlined in Section~\ref{methods}.   We see substantial heterogeneity between neighborhoods in Philadelphia, both in terms of the their estimated crime levels ($\hat{\alpha}_i$'s) and changes in crime over time ($\hat{\beta}_i$'s).  Regardless of the model, most neighborhoods in the city show decreasing trends in crime over time (negative $\beta$'s) with a small subset of neighborhoods showing an increasing trend.

The shrinkage imposed by the global shrinkage model is more visually striking for the change in violent crime over time than the overall level of crime.  The maps of the $\hat{\alpha}_i$'s from the no shrinkage and global shrinkage models are almost indistinguishable in Figure~\ref{fig-alpha} whereas the map of the $\hat{\beta}_i$'s from the global shrinkage model has been shifted substantially compared to the no shrinkage map in Figure~\ref{fig-beta}.    This observation suggests that there is more substantial heterogeneity between neighborhoods in terms of their overall level of crime compared to their change in crime over time.  

This heterogeneity in the mean level of crime is expected as it is influenced by many years of transformation in the city of Philadelphia that led to its current built and social environment.   Differences in these overall spatial crime patterns can be addressed by urban planners, whose effects are long-lasting \citep{johnson2008stable}.  In contrast, differences in the trend over time identify shorter-term patterns, which can be addressed with interventions by local police departments.  

The overall level of crime also seems to have a greater inherent spatial correlation between proximal neighborhoods than the change in crime over time.   The Moran's I values calculated from the estimated $\hat{\alpha}_i$'s are I = 0.33 for both the no shrinkage and global shrinkage models, compared to the value of I = 0.17 from the estimated $\hat{\beta}_i$'s for those same models in Table~\ref{tb:cv}.  This is clear also from the maps from the no shrinkage model (top left) in Figures~\ref{fig-alpha} and \ref{fig-beta}: the estimated $\hat{\beta}_i$'s are more ``spotty'' and less smooth than the corresponding map of the $\hat{\alpha}_i$'s.

However, once we build spatial correlation into our model via the spatial CAR prior (\ref{eq:joint_leroux1})-(\ref{eq:joint_leroux2}), the resulting $\hat{\beta}_i$'s are more spatially correlated than the resulting $\hat{\alpha}_i$'s, as can be seen in the lower left of Figures~\ref{fig-alpha} and Figures~\ref{fig-beta} as well as the corresponding Moran's I = 0.53 for the $\hat{\alpha}_i$'s versus I = 0.61 for the $\hat{\beta}_i$'s.  Note that all these reported Moran's I values have a standard error approximately equal to 0.016, and so they are all significantly different from the null hypothesis of no spatial autocorrelation. 

Although the smoother maps from the spatial CAR model (lower left of Figures~\ref{fig-alpha} and ~\ref{fig-beta}) ease  interpretation by identifying larger regions of the city with similar crime dynamics, there is the potential to over-shrink certain neighborhoods that should actually stand out from their neighbors.  In any large city, natural or artificial {\it barriers} such as rivers, highways or rail lines create discontinuities between neighborhoods which should not be smoothed over.  In Section~\ref{results-borders}, we examine the results from our model extension that allows a subset of borders between neighborhoods to be turned into barriers.

\subsection{Borders turned into Barriers} \label{results-borders}

In Section~\ref{varW}, we extended the spatial CAR model to allow a subset of the weights $w_{ij}$ to vary, which allows the {\it borders} ($w_{ij} = 1$) between some neighborhoods to be changed into {\it barriers} ($w_{ij} = 0$); the latter prevent shrinkage between two bordering neighborhoods.  Our model has separate weight matrices $\W^\alpha$ and $\W^\beta$, so a particular border can be turned into a barrier either for the level of crime ($\alpha_i$'s) or the change in crime over time ($\beta_i$'s) or both.  Using this model, we estimate the posterior probability that we change a border into a barrier for each border between proximal neighborhoods in Philadelphia. 

\begin{figure}
\includegraphics[width = \textwidth]{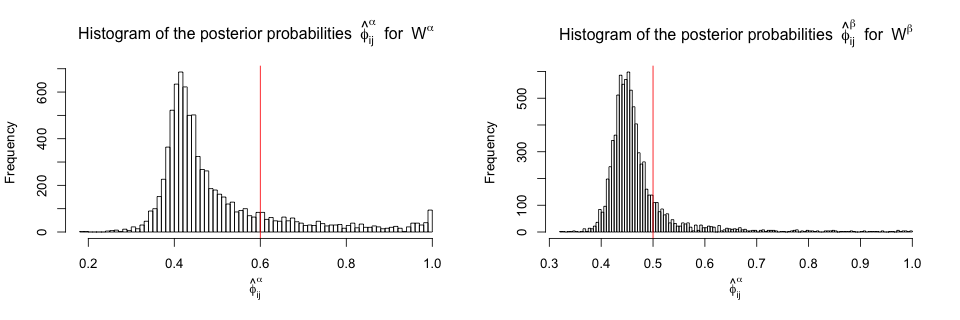} %
\caption{Histograms of the posterior probabilities of each border being turned into a barrier. {\bf Left:} Probabilities for barriers for the $\alpha_i$'s; the threshold to identify the borders turned into barriers is 0.6 (red line). {\bf Right:} Probabilities for barriers for the $\beta_i$'s; the threshold to identify the borders turned into barriers is 0.5 (red line).} \label{fig-hist-prob}
\end{figure}

Figure~\ref{fig-hist-prob} gives the distribution of the estimated posterior probability of a border being turned into barrier for each border encoded in the weight matrices $\W^\alpha$ and $\W^\beta$.  These distributions seem to have two components: a main mode representing the behavior of the majority of the borders, which has a low probability of being turned into a barrier, and a ``tail'' component which has a higher probability of being turned into a border. 

It is clear that many more borders have a high probability of being a barrier for the level of crime ($\alpha_i$'s) compared to the change in crime over time ($\beta_i$'s).  In other words, our variable border model is detecting more discontinuities between bordering neighborhoods in the level of crime compared to the change in crime over time.  In Section 5 of our Supplementary Materials \citep{supplement}, we explore an alternative model that only allows variable borders for the mean level of crime.  

In the lower right panels of Figures~\ref{fig-alpha} and ~\ref{fig-beta}, we provide maps of Philadelphia where we have highlighted any borders between neighborhoods that have been inferred by our model to have a high probability of being barriers.  These particular highlights are based on posterior probabilities larger than $60\%$ for $\W^\alpha$ and larger than 50\% for $\W^\beta$.  

We see in the lower right panel of Figure \ref{fig-alpha} that barriers have been detected around several parks including Fairmount Park, Wissahickon Valley Park, and Pennypack Creek Park (indicated by the black numbers 1, 2 and 3 respectively in the lower right panel of Figure \ref{fig-alpha}).
In these cases, our model has automatically detected several natural geographic structures within Philadelphia as locations which have discontinuities in the level of crime.    

We also see that some estimated barriers have isolated particular neighborhoods from their proximal neighbors.  For example, the neighborhood of Bridesburg (indicated by the black number 4 in the lower right panel of Figure \ref{fig-beta}) seems to have a much more positive trend on crime over time than its surrounding neighborhoods.  

As barriers highlight the boundaries of regions that display differences in either in the level of crime or the trend in crime over time, these barriers can be used by police departments and city planners for delineating the possible limits of effectiveness for interventions or as potential targets for interventions themselves.   

\subsection{Neighborhoods with Most Extreme Crime Trends} \label{results-bestworst}

To further understand which regions of Philadelphia have the most extreme levels of crime and trends in crime over time, we can examine the most extreme intercepts ($\alpha_i$'s) and slopes ($\beta_i$'s) found by our fitted models.  Specifically, we focus on the estimated $\alpha_i$'s and $\beta_i$'s from the local shrinkage spatial CAR model (\ref{eq:joint_leroux1})-(\ref{eq:joint_leroux2}) that had the best out-of-sample predictive performance in Table~\ref{tb:cv}.

\begin{figure}
\includegraphics[width = 4in]{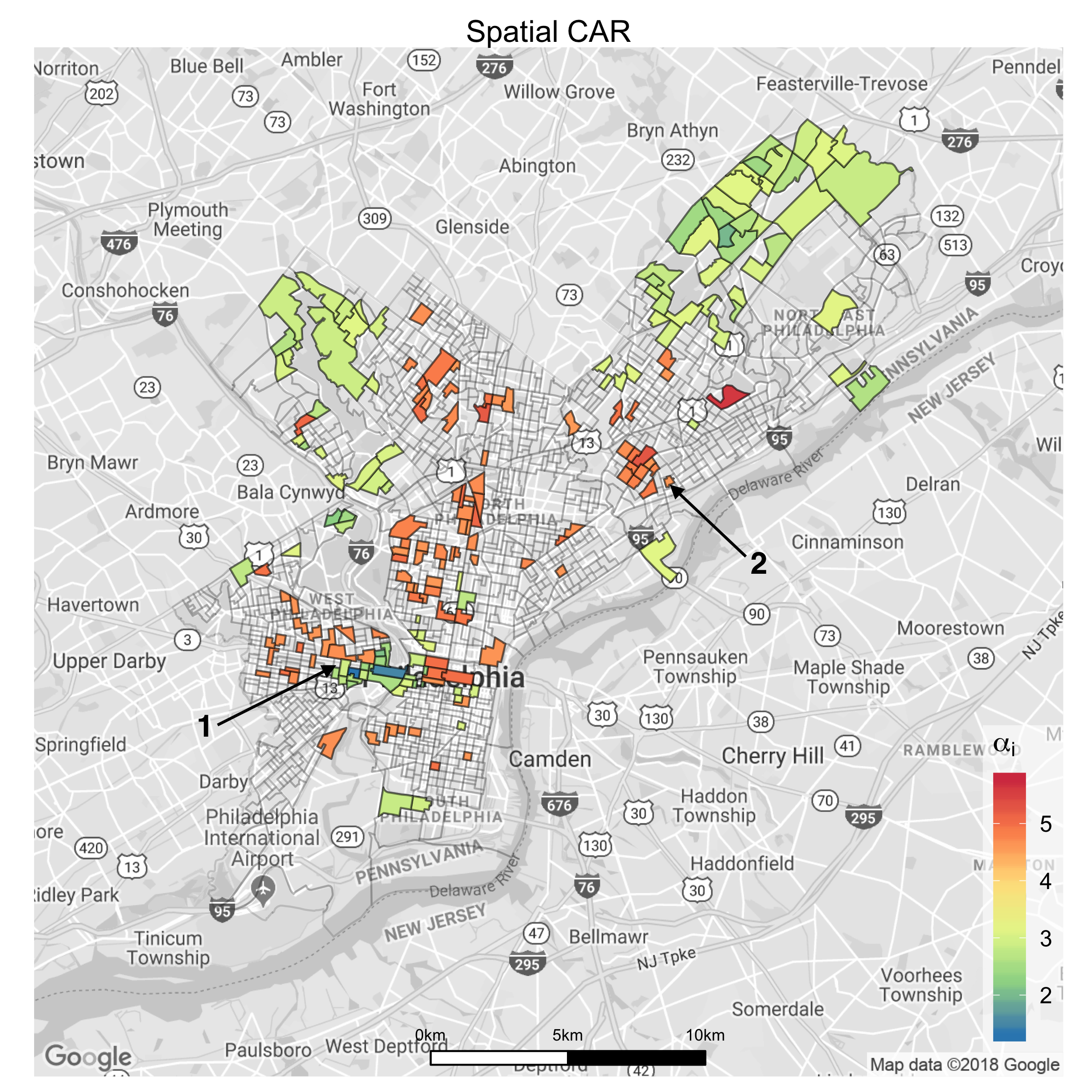} %[scale = 0.16] %[width = \textwidth]
\includegraphics[width = 4in]{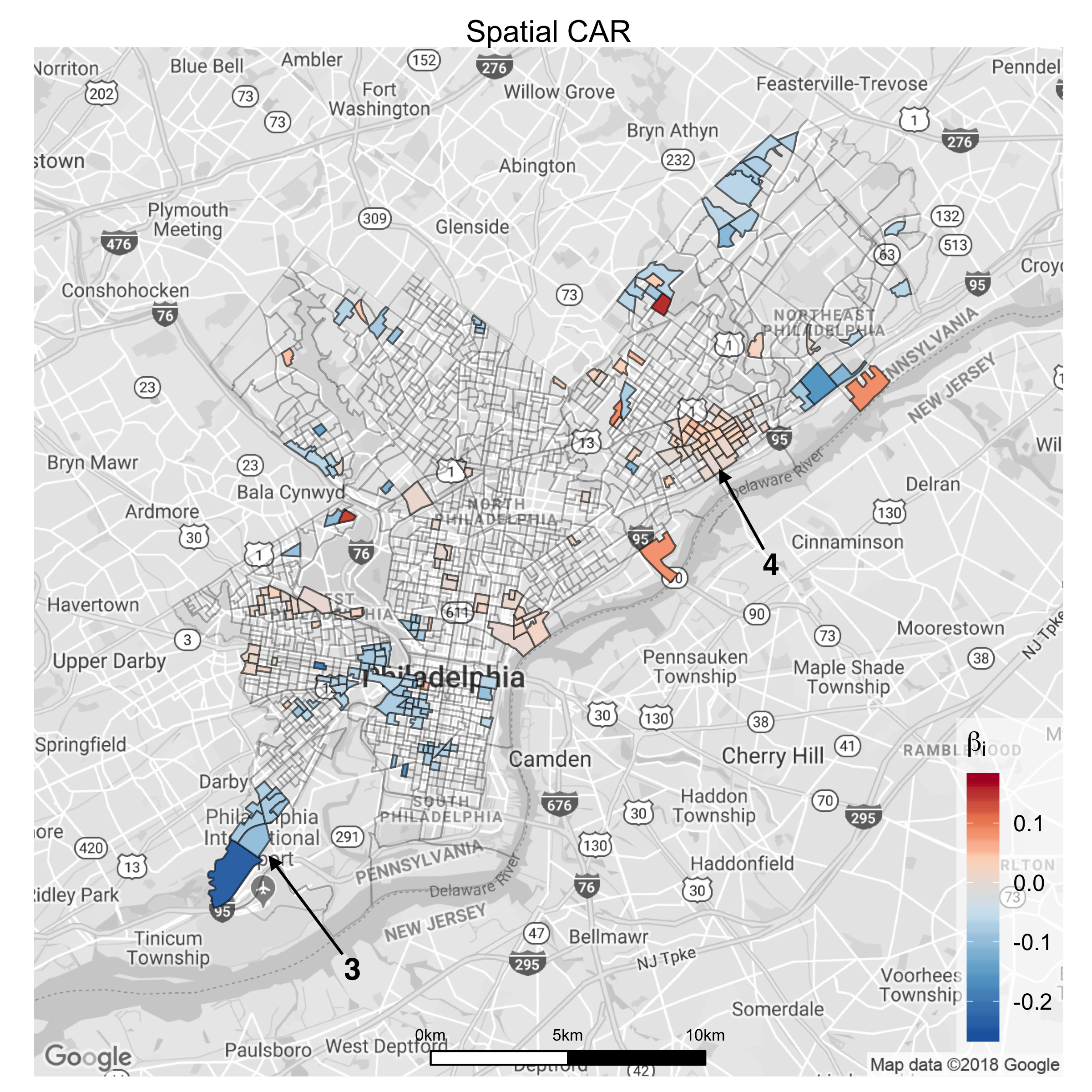} %[scale = 0.16] %[width = \textwidth]
\caption{{\bf Top:} The 50 neighborhoods with the largest $\hat{\alpha}_i$'s (red) and 50 neighborhoods with the smallest $\hat{\alpha}_i$'s (green). {\bf Bottom:} The 50 neighborhoods with the largest $\hat{\beta}_i$'s and 50 neighborhoods with the smallest $\hat{\beta}_i$'s. These maps were created with the R package {\tt ggmap} \citep{kahle2013ggmap}.}
\label{fig-extreme}
\end{figure}

Figure~\ref{fig-extreme} provides maps that highlight the most extreme (largest 50 and smallest 50) neighborhoods in terms of the estimated level of crime ($\hat{\alpha}_i$'s) and in terms of the estimated change in crime over time ($\hat{\beta}_i$'s).  

We see that the region of University City in West Philadelphia (black number 1 in the top panel of Figure \ref{fig-extreme}) is an interesting transitional area that contains both neighborhoods with the highest and lowest levels of crime in the city.  We also see that the area of Frankford (black number 2 in the top panel of Figure \ref{fig-extreme}) has neighborhoods with high levels of crime.  This area is a major transportation hub for the Northeast region of Philadelphia.

The SW region of Philadelphia, specifically the Elmwood and Eastwick neighborhoods (black number 3 in the bottom panel of Figure \ref{fig-extreme}) have seen some of the largest reductions in crime over the past decade in Philadelphia.  We also see some regions of the city that are showing increases in crime over that same time period, such as the Wissinoming and Tacony neighborhoods (black number 4 in the bottom panel of Figure \ref{fig-extreme}) that are just to the northeast of the high crime neighborhoods of Frankford (black number 2 in the top panel of Figure \ref{fig-extreme}).  

In Section 5 of our Supplementary materials \citep{supplement}, we provide additional visualizations of the neighborhood-specific parameters that are significantly different from the overall mean across the city as well as the widths of the credible intervals for these parameters.   

\section{Discussion} \label{discussion}

Reliable estimation of the change in crime over time at the local neighborhood level is a crucial step towards a better understanding of the determinants of public safety in large urban areas.  With a focus on the city of Philadelphia, we have explored several Bayesian approaches to modeling crime trends within the areal units of neighborhoods while sharing information either globally or locally across the city.  

Imposing local shrinkage between proximal neighborhoods via a spatial conditional autoregressive (CAR) prior gives the best out-of-sample predictions of violent crime compared to models that impose global shrinkage or no shrinkage at all between neighborhoods.  We also explore allowing the weight matrix of our spatial CAR model to vary in order to detect neighborhood borders that represent spatial discontinuities in the level of crime or change in crime over time.  In this way, we automatically detect several natural barriers in the geography of Philadelphia.   Our model estimates also identify the regions of Philadelphia with the most extreme levels of violent crime as well as the largest increases and reductions in crime over the period of 2006-2015.

\section{Acknowledgements} 

We thank Rachel Thurston and Theresa Smith for helpful contributions to our modeling efforts and interpretation of our results. 

\begin{supplement}[id=suppA]
  %\sname{Supplement A}
  \stitle{Supplementary materials for ``Spatial modeling of trends in crime over in Philadelphia"}
  \slink[doi]{COMPLETED BY THE TYPESETTER}
  \sdatatype{.pdf}
  \sdescription{We provide maps outlining the block group structure, population count, and distribution of violent crimes in Philadelphia.  We give implementation details for our fitted models.  We compare results under an alternative choice of prior distributions and under an alternative model that only allows variable borders for the mean level of crime. We provide additional numerical details about our estimated partial effects.  We also provide visualizations of the significance and widths of credible intervals for the neighborhood-specific parameters.}
\end{supplement}

\bibliography{urban.spatial.crime.arxiv}
\bibliographystyle{imsart-nameyear}

%\widetext
\newpage
\begin{center}
{\Large {\bf Supplementary Materials for }}

\bigskip

{\Large {\bf ``Spatial Modeling of Trends in Crime over Time in Philadelphia"}}

\bigskip

\end{center}
%%%%%%%%%% Merge with supplemental materials %%%%%%%%%%
%%%%%%%%%% Prefix a "S" to all equations, figures, tables and reset the counter %%%%%%%%%%
\setcounter{equation}{0}
\setcounter{figure}{0}
\setcounter{table}{0}
\makeatletter
\renewcommand{\theequation}{S\arabic{equation}}
\renewcommand{\thefigure}{S\arabic{figure}}
\renewcommand{\bibnumfmt}[1]{[S#1]}
\renewcommand{\citenumfont}[1]{S#1}
%%%%%%%%%% Prefix a "S" to all equations, figures, tables and reset the counter %%%%%%%%%%

\section{Maps of Data in Philadelphia}

\begin{figure}[ht!]
\renewcommand\thefigure{S1}
\centering
\includegraphics[width=2.5in]{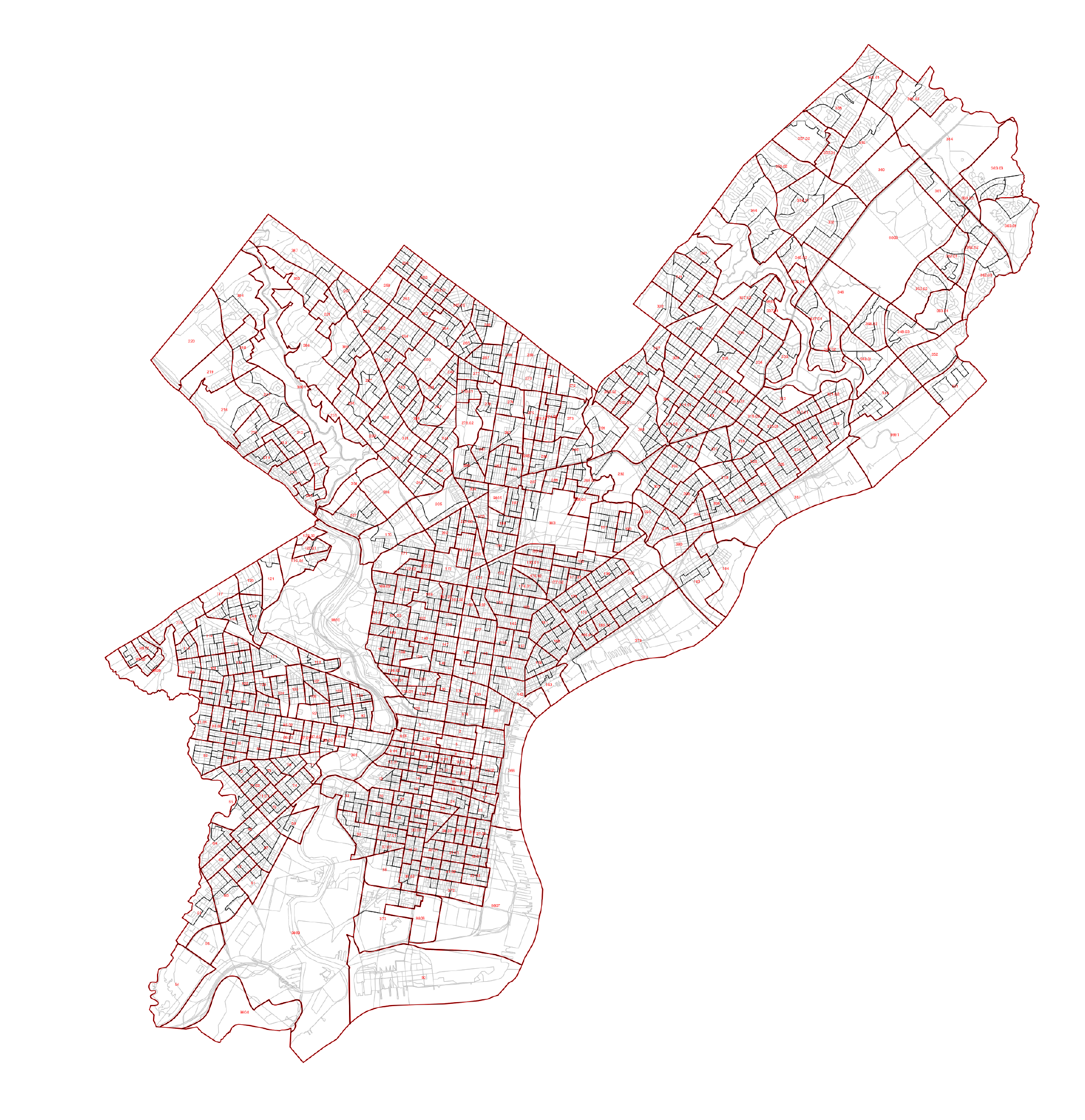}~
\includegraphics[width=2.5in]{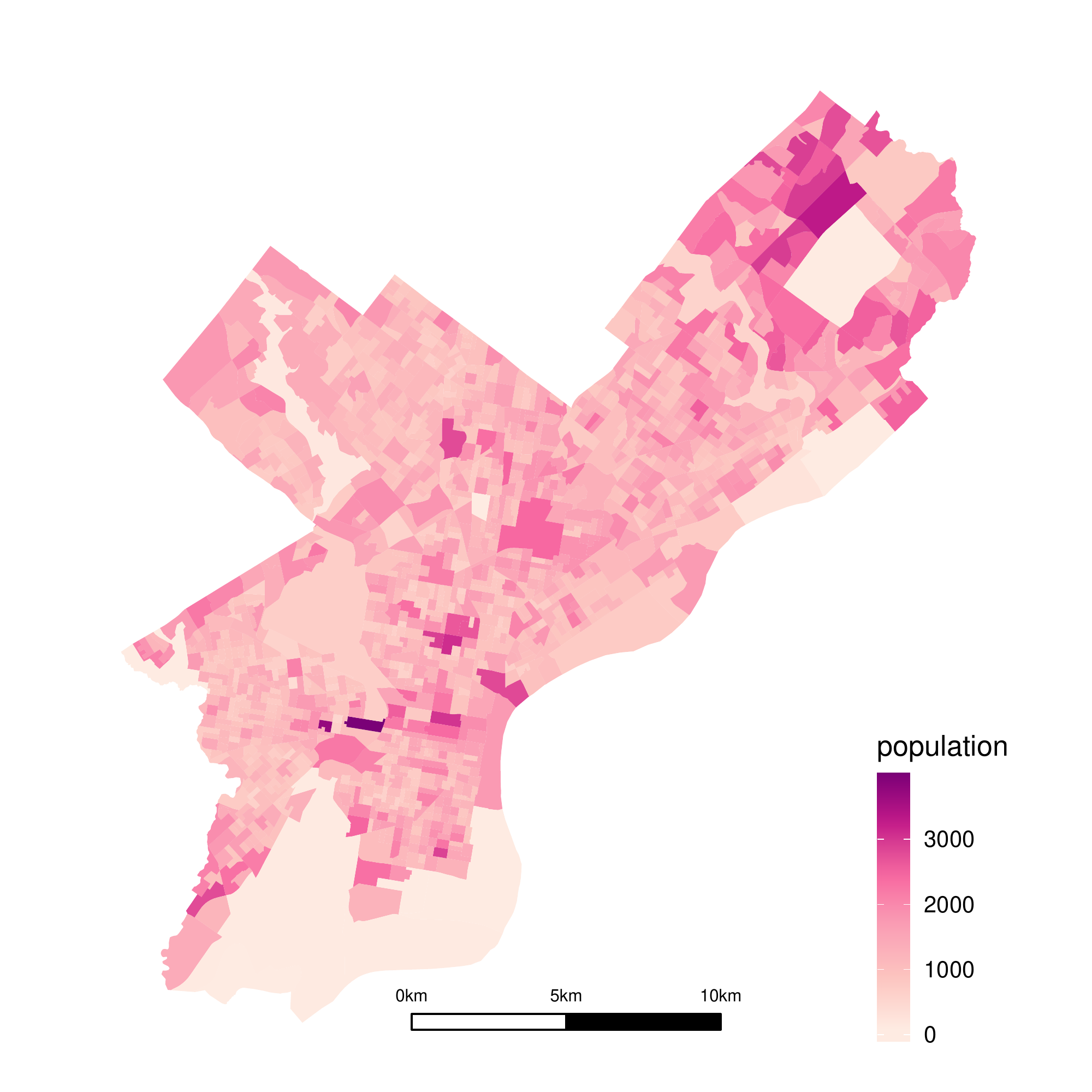}

\caption{{\bf Left:} Map of Philadelphia divided into census tracts (red lines) and block groups (black lines) by US Census Bureau.  {\bf Right:} Population count by block group in Philadelphia. These maps were created with the R package {\tt ggmap} \citep{kahle2013ggmap}.}
\label{map-philadelphia-blockgroups}
\end{figure}

Figure~\ref{map-philadelphia-blockgroups} (left) gives a map outlining the 1336 block groups in Philadelphia.    Figure~\ref{map-philadelphia-blockgroups} (right) shows population count for each block group in Philadelphia.  

\begin{figure}[ht!]
\renewcommand\thefigure{S2}
    \centering
    \begin{subfigure}[t]{0.5\textwidth}
        \centering
        \includegraphics[width=\textwidth]{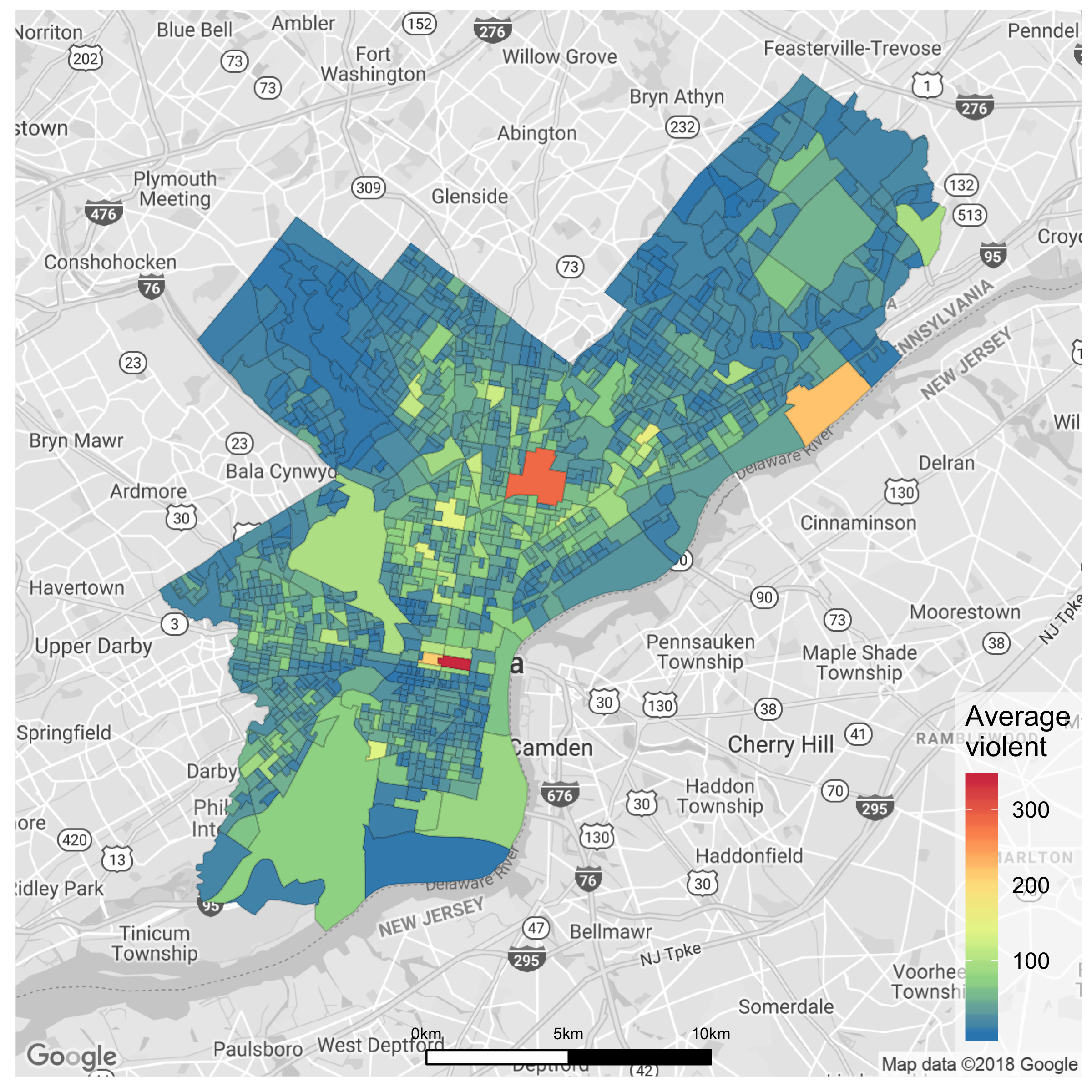}
    \end{subfigure}%
    ~ 
    \begin{subfigure}[t]{0.5\textwidth}
        \centering
        \includegraphics[width=\textwidth]{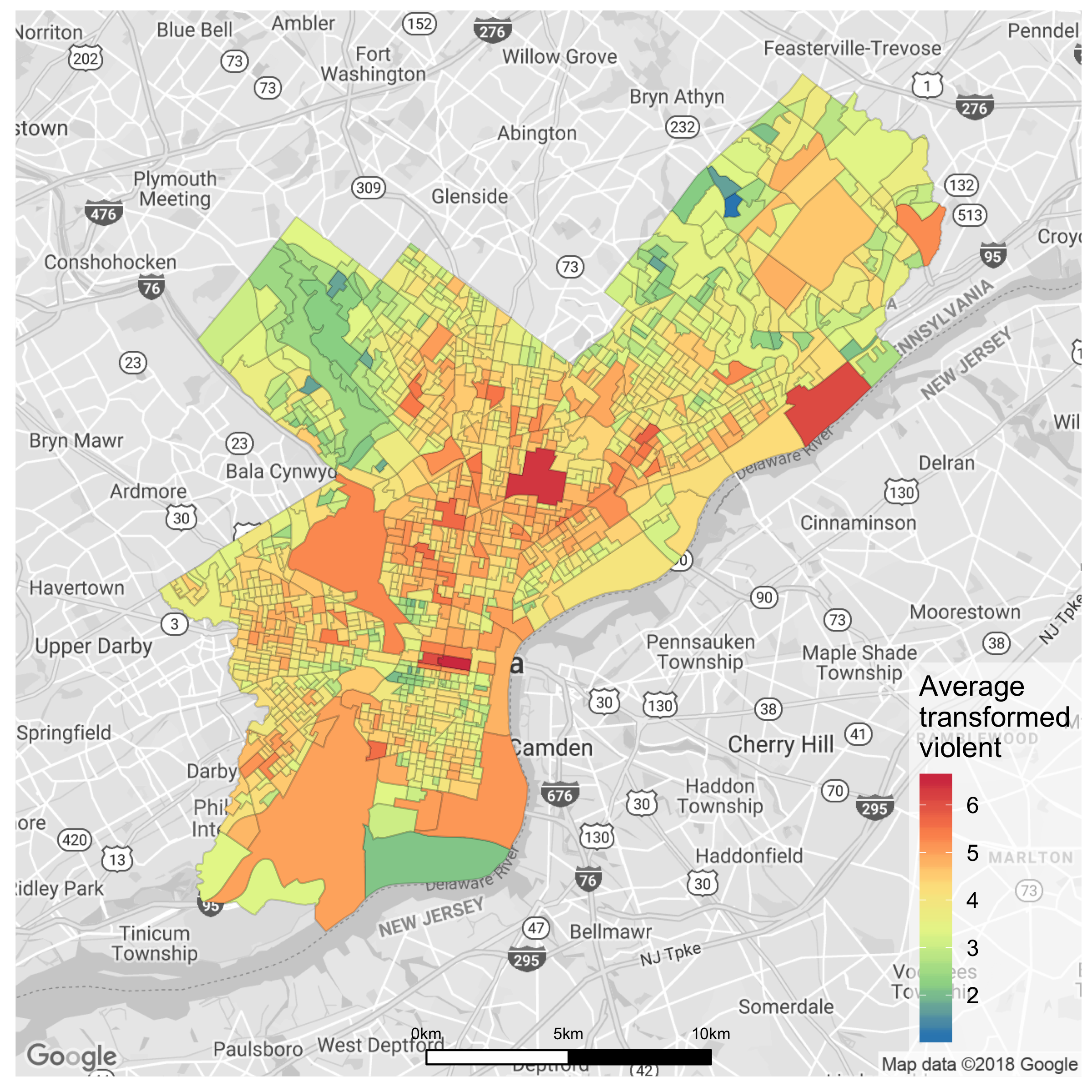}
    \end{subfigure}
    \caption{Distribution of violent crime over the block groups of Philadelphia. {\bf Left:} violent crimes per block group, averaged over the years from 2006 to 2015. {\bf Right:} logarithm of violent crimes per block group, averaged over the years from 2006 to 2015. These maps were created with the R package {\tt ggmap} \citep{kahle2013ggmap}.} \label{crime-spatial}
\end{figure}

In Figure~\ref{crime-spatial} (left), we give the count of violent crimes per year in each block group, averaged over the years 2006-2015.  We see substantial heterogeneity across block groups in the average counts of violent crimes per year.   There are several outlying values: particular block groups that have much higher average violent crime counts.  The largest among these is the Market East neighborhood in central Philadelphia.  

These outlying neighborhoods motivate us to examine violent crime totals on the log scale.   In Figure~\ref{crime-spatial} (right), we give the average of the logarithm of the count of violent crimes per year in each block group, averaged over the years 2006-2015.  We can see more details of the spatial distribution of violent crime on the log scale.  Modeling crime on the log scale has the additional benefit that changes in log crime can be interpreted as percentage changes in crime. 

In both Figure~\ref{crime-spatial} (left) and (right), we see evidence of spatial correlation in violent crime totals between proximal block groups throughout the city. 

\section{Gibbs sampling}

In section 3 of our paper, we described the different models considered in this work; here we illustrate the Gibbs sampling strategy to sample from their posterior distributions. We are mainly interested in the coefficients $\Bgamma$ for the predictors and
in the collection of neighborhood-specific coefficients $(\Balpha, \Bbeta)$.  $\Bgamma$ have a Normal prior distribution with mean zero and covariance matrix proportional to the identity matrix. $\Balpha$ and $\Bbeta$ also have Normal prior distributions with mean zero, but have different covariance matrices depending on the model.  In the global shrinkage model, the covariance matrices are proportional to the identity matrix, while in the local shrinkage CAR model the covariance matrices depend on the Laplacian matrix of the geography. 

We denote with $\Y$ the $N$-dimensional vector (where $N = nT$) found by concatenating the $y_{it}$'s, ordered by block group: $\Y = (y_{11}, \ldots, y_{1T}, \ldots, y_{n1},$ $ \ldots, y_{nT})$; moreover let $\Btheta = (\Bgamma, \Balpha, \Bbeta)$ represent the collection of the coefficients. Let the matrix $\X$ be designed so that the covariates corresponding to block group $i$ at time $t$ are contained in row $(i-1)T + t$ and by multiplying this row with the vector of parameters we get  $X_{(i-1)T + t}^\gamma \Btheta  = \sum_j \gamma_j z_{ij} + \alpha_i + \beta_{i}t$.  
With this notation, the conditional distribution of the data is given by $\Y \vert \Btheta,\sigma^2 \sim N(\X\Btheta, \sigma^2 \I )$. 

We can also combine the prior distributions of $\Bgamma, \Balpha, \Bbeta$ to get the distribution of $\Btheta$: let $\Btheta_0 = (\0, \alpha_0 \1, \beta_0 \1)$ a ($d + 2n$)-dimensional vector representing the conditional mean of $\Btheta$ and let $\BOmega_0=\BSigma_0^{-1}$ be the block matrix representing its precision matrix. 
Since $p(\Bgamma) = {\rm N}(\0, \tau_\gamma^2 \cdot \I)$, the first $d \times d$ diagonal block of $\BOmega_0$ is equal to $\tau_\gamma^{-2} \I$; the next two $n \times n$ diagonal blocks instead are the precision matrices of $\Balpha$ and $\Bbeta$: $\tau_\alpha^{-2} \BSigma^{-1}$ and $\tau_\beta^{-2} \BSigma^{-1}$, where $\BSigma = \I$ in the global shrinkage model and $\BSigma^{-1}= [\rho(\D_W-\W) +(1-\rho)\I] $ in the spatial CAR model; the remaing blocks are zero matrices. 
Then $p(\Btheta \vert \Btheta_0, \tau_\alpha^2,\tau_\beta^2) \propto \exp \left( -\frac12 (\Btheta - \Btheta_0)^T\BOmega_0(\Btheta - \Btheta_0) \right)$.
Moreover, we set non-informative flat priors on $\alpha_0, \beta_0$, and the variance hyper-parameters $\sigma^2$, $\tau_\gamma^2$, $\tau_\alpha^2$ and $\tau_\beta^2$ have Inverse-Gamma priors, tuned in an Empirical Bayes fashion.

%% This paragraph was not changed
\paragraph{Posterior conditional distribution of $\Btheta$}
With this notation, we can find the conditional posterior distribution of $\Btheta$ as we would do in a usual linear regression:
$$
\Btheta | \Y, \Btheta_0,\sigma^2, \Btau^2  \sim N(\hat{\Btheta}, \V_{\Btheta})
$$
where 
\begin{align*}
\hat{\Btheta} &= \left(\BOmega_0 + \X^T\X / \sigma^2 \right)^{-1} (\BOmega_0 \Btheta_0+ \X^T \Y / \sigma^2) \\
\V_{\Btheta} &= \left( \BOmega_0+ \X^T\X / \sigma^2 \right)^{-1}.
\end{align*}

%% This paragraph was not changed
\paragraph{Posterior conditional distribution of $\Btheta_0$}
Similarly, the posterior distribution of the mean hyper-parameters $\alpha_0$ and $\beta_0$ can be found as
\begin{align*}
\alpha_0 \vert \Balpha, \tau_\alpha^2 &\sim N\left(\frac{\1^T \Sigma_\alpha^{-1}\Balpha}{\1^T \Sigma_\alpha^{-1}\1},  \frac{\tau_\alpha^2}{\1^T \Sigma_\alpha^{-1}\1}\right)\\
\beta_0 \vert \Bbeta, \tau_\beta^2 &\sim N\left(\frac{\1^T \Sigma_\beta^{-1}\Bbeta}{\1^T \Sigma_\beta^{-1}\1},  \frac{\tau_\beta^2}{\1^T \Sigma_\beta^{-1}\1}\right).
\end{align*}

%% This paragraph WAS changed!
\paragraph{Posterior conditional distribution of $\sigma^2$, $\tau^2_\gamma$, $\tau^2_\alpha$ and $\tau^2_\beta$} 
For the variance hyper-parameters $\sigma^2, \tau^2_\gamma, \tau_\alpha^2$ and $\tau_\beta^2$, the prior distributions are 
\begin{align*}
\sigma^2 &\sim \mbox{Inv-Gamma}(a_\sigma, b_\sigma)\\
\tau^2_\alpha &\sim \mbox{Inv-Gamma}(a_\alpha, b_\alpha)\\
\tau^2_\beta &\sim \mbox{Inv-Gamma}(a_\beta, b_\beta)\\
\tau^2_\gamma &\sim \mbox{Inv-Gamma}(a_\gamma, b_\gamma).
\end{align*}
where the hyper-parameters are tuned in an empirical Bayes fashion so that the prior mean of the variance parameters is equal to the variance estimated from the model with no shrinkage, and the prior variance is small.
The conditional posterior distributions are also Inverse-Gamma:
\begin{align*}
\sigma^2 | \Y, \Btheta &\sim \mbox{Inv-Gamma}\left(a_\sigma + \frac{N}{2}, b_\sigma+\frac12\sum_{i=1}^n\sum_{t=1}^T (y_{it} -\z_i^T\Bgamma - \alpha_i - t\beta_i)^2 \right)\\
\tau_\gamma^2 \vert \Y, \Bgamma &\sim \mbox{Inv-Gamma}\left(a_\gamma + d/2, b_\gamma + \frac12\sum_{j=1}^d\gamma_{j}^2 \right)\\
\tau^2_\alpha | \Y,  \Balpha, \alpha_0 &\sim \mbox{Inv-Gamma}\left(a_\alpha+\frac{n}{2}, b_\alpha + (\Balpha-\alpha_0\1)^T \BSigma_\alpha^{-1} (\Balpha-\alpha_0\1)/2 \right)\\
\tau^2_\beta | \Y, \Bbeta, \beta_0 &\sim \mbox{Inv-Gamma}\left(a_\beta+\frac{n}{2}, b_\beta+(\Bbeta-\beta_0\1)^T \BSigma_\beta^{-1} (\Bbeta-\beta_0\1)/2 \right).
\end{align*}

\paragraph{Posterior conditional distribution of $\rho$}
The prior distribution on $\rho$ is Beta$(10,$ $10)$, and since its conditional posterior distribution does not have a closed form, we sample this parameter with a Metropolis Hasting procedure. Given its past value $\rho^{t-1}$ we propose a new candidate $\rho^*$ with density $g(\rho^* \vert \rho^{t-1}) = {\rm Beta}(b \rho^{t-1}/(1-\rho^{t-1}), b)$; this parametrization allows the mean to be $\rho^{t-1}$ and the variance to be small when we choose $b = 10$. The acceptance probability is then 
$$
a = 1 \wedge \frac{p(\rho^*\vert e.e.)}{p(\rho^{t-1}\vert e.e.)}\frac{g(\rho^{t-1} \vert \rho^*)}{g(\rho^* \vert \rho^{t-1})}
$$
where the posterior conditional distribution $p(\rho \vert e.e.)$ is proportional to the product of the prior distribution of $(\Balpha,\Bbeta)$ given $\rho$ and the prior of $\rho$.  We use the notation $e.e.$ to denote ``everything else", i.e. the current values of all other parameters in the model.

\paragraph{Posterior conditional distribution of $\W$} 

Finally, in model 
(3.13)-(3.14)
we allow the adjacency matrix itself to be random. We consider all the pairs of regions that share a border ($w_{ij} = 1$) and we allow those borders to potentially become barriers ($w_{ij} = 0$). We model these variable weights as $w_{ij}^\alpha \vert \phi^\alpha \sim {\rm Bern}(\phi^\alpha)$ and independently, $w_{ij}^\beta \vert \phi^\beta \sim {\rm Bern}(\phi^\beta)$, with $\phi^\alpha, \phi^\beta \overset{iid}{\sim} {\rm Beta}(1,9)$.  

With these prior distributions, the conditional posterior distribution for $\W^\alpha$ is 
\begin{align*}
&p(\W^\alpha \vert e.e.) \propto p(\Balpha| \alpha_0,\tau_\alpha^2,\rho,\W^\alpha)p(\W^\alpha \vert \phi^\alpha) \\
& \propto  \det(\BSigma_\alpha^{-1})^{1/2}  \exp \left( -\frac{1}{2\tau_\alpha^2}(\Balpha-\alpha_0\1)^T \BSigma_\alpha^{-1}(\Balpha-\alpha_0\1) \right)p(\W^\alpha \vert \phi^\alpha)\\
& \propto  \det(\BSigma_\alpha^{-1})^{1/2}  \exp \left( -\frac{\rho}{2\tau_\alpha^2}(\Balpha-\alpha_0\1)^T  (\D_{W^\alpha} - \W^\alpha)(\Balpha-\alpha_0\1) \right)p(\W^\alpha \vert \phi^\alpha).
\end{align*}
Note that, because of the determinant term, the entries of $\W^\alpha$ are not independent {\it a posteriori}. Thus we sample each entry $w_{ij}^\alpha=w_{ji}^\alpha$ conditional on the rest of the matrix $\W^\alpha_{-ij}$ as $p(w_{ij}^\alpha =1 \vert e.e.) = q$, where
%. And while the dependency of the determinant term on $w_{ij}^\alpha$ is not obtainable in closed form, we can simplify the dependency on $w_{ij}$ in the exponential term:
%\frac{p(w_{ij}^\alpha = 1 \vert e.e.)}{p(w_{ij}^\alpha = 0 \vert e.e.)}&
\begin{align*}
\frac{q}{1-q} = \sqrt{\frac{\det(\BSigma_\alpha^{-1}(w_{ij}^\alpha = 1))}{\det(\BSigma_\alpha^{-1}(w_{ij}^\alpha = 0))}} \exp\left( -\frac{\rho}{2\tau_\alpha^2} (\alpha_i - \alpha_j)^2\right) \frac{\phi^\alpha}{1-\phi^\alpha}.
\end{align*}
A highly similar procedure (with the obvious substitutions) is used to sample the entries of $\W^\beta$. 

\paragraph{Posterior conditional distribution of $\phi$} To express the prior information that only a small percentage of the borders should be turned into barrier, the prior distribution of $\phi^\alpha$ and $\phi^\beta$ is Beta$(1,9)$. 
Since the $w_{ij}^\alpha$ and $w_{ij}^\beta$ are Bernoulli distributed, the posterior distributions for $\phi^\alpha$ and $\phi^\beta$ are
%Their conditional posterior distribution can be easily found because of the conjugate model: 
\begin{align*}
\phi^\alpha \vert \W^\alpha &\sim {\rm Beta} \left( 1 +\sum_{(i,j) \in I} w_{ij}^\alpha,9 +\sum_{(i,j) \in I} (1-w_{ij}^\alpha)\right)\\
\phi^\beta \vert \W^\beta &\sim {\rm Beta} \left( 1 +\sum_{(i,j) \in I} w_{ij}^\beta,9 +\sum_{(i,j) \in I} (1-w_{ij}^\beta)\right).
\end{align*}

\section{MCMC Implementation Details}

The results reported in Table 1 of our paper for the hierarchical models have been implemented using Gibbs sampling. In particular, for each model 1000 samples where used, after discarding a burn-in period of 50 iterations and thinning every 2 samples. By running multiple chains and superimposing their trace plots, we noted that the convergence happened after a relatively short time and that samples were not highly correlated.

\section{Prior Robustness for Variance Hyperparameters} \label{noninf}

In Section~3 of our main paper, we used priors for the variance parameters with hyper-parameters that were tuned in an Empirical Bayes fashion.  In this section, we show that highly similar results are obtained when using more non-informative prior distributions on these variance parameters.  

In particular, we consider a uniform prior on $\Bgamma, \log \sigma, \tau_\alpha$ and $\tau_\beta$, which is equivalent to:
\begin{align*}
p( \Bgamma)&\propto1\\
p( \sigma^2)&\propto \sigma^{-2}\\
p( \tau_\alpha^2)&\propto \tau_\alpha^{-1}\\
p( \tau_\beta^2)&\propto \tau_\beta^{-1}.
\end{align*}
%$$p( \Bgamma, \log \sigma, \tau_\alpha,\tau_\beta) \propto1.$$

Table~\ref{tb:noninf} is equivalent to Table 1 in our main paper but with results from the estimated models that use the non-informative priors given above.   Only the global, spatial CAR and variable border model results are reported since priors are not involved in the no-shrinkage model (3.5).   

\begin{table}[ht!]
\renewcommand\thetable{S1}
\centering
\begin{tabular}{lccccc}
Model  & ${\rm MSE}_{\rm in}$ & ${\rm MSE}_{\rm out}$  & ${\rm MSE}_{\rm cv}$& Moran's $I$\\ 
\hline
Separate $\alpha_i \, , \, \beta_i$ Models & & & & \\
\phantom{y} Global Shrinkage & 0.0698 & 0.1080  & 0.0927 & 0.17 \\ 
\phantom{y} Spatial CAR   &  0.0701 & 0.1052  & 0.0922  & 0.61\\
\phantom{y} Variable Borders & 0.0706 & 0.1069  & 0.0927 & 0.48 \\
%\phantom{y} Variable Borders - only A & 0.0709 & 0.1049  & 0.0920 & 0.64 
\end{tabular} 
\caption{Comparison of predictive accuracy between the different models outlined in Section~3 with non-informative priors on the hyper parameters.  The mean squared error for both in-sample and out-of-sample predictions are provided.  We also provide the Moran's $I$ measure of spatial correlation calculated on the estimated time trends $\beta_i$ from each model.\label{tb:noninf}}
\end{table}

Examining Table~\ref{tb:noninf}, we see almost the exact same predictive results as the predictive results given in Table 1 of our main paper.  There are very slight numerical differences in the Spatial CAR in-sample error, the Global Shrinkage out-of-sample error and the Variable Borders Moran's I, but these differences could easily be attributed to MCMC sampling variability.

\section{Additional Model Results} \label{extramodelresults}

In Section~\ref{numres} we report the numerical estimates of the partial effects, which are shown in Figure~2 of the main paper. In Section~\ref{randomWalpha} we describe the results from a model with variable borders for $\alpha_i$ but fixed borders for $\beta_i$. In section~\ref{variability} we provide different visualizations of the variability in the estimated neighborhood-specific coefficients $\alpha_i$ and $\beta_i$.

\subsection{Numerical results for partial effects\label{numres}}
In table~\ref{tb:covariates} we report the numerical values corresponding to the partial effects shown in Figure 2 of the main paper. For each model with neighborhood-specific coefficients outlined in Section 3 of the paper, we report maximum likelihood estimates, standard errors, posterior means and posterior standard deviations.

\begin{table}[ht!]
\renewcommand\thetable{S2}
\centering
\begin{tabular}{lrrrrrrrr}
\hline
  & \multicolumn{2}{c}{No shrinkage} & \multicolumn{2}{c}{Global shrinkage}& \multicolumn{2}{c}{Spatial CAR}& \multicolumn{2}{c}{Variable borders} \\ 
\hline
  & Estimate & St.Error & Mean & SD & Mean & SD & Mean & SD\\ 
\hline
log.income & -0.186 & 0.009 &-0.186   & 0.022 & -0.113 &   0.023&-0.099 &  0.021 \\
sqrt.poverty  & 0.182 &0.009 & 0.182   &0.023 &0.095  &  0.020 & 0.110 &  0.018\\ 
segregation  &0.010 & 0.005  & 0.010   &0.013& -0.023 &  0.017 & -0.013  & 0.017\\ 
sqrt.vacantprop  &0.116  & 0.006 & 0.115  & 0.014& 0.052  &  0.015  &  0.045 &  0.014\\
sqrt.comresprop  & 0.227 & 0.005  & 0.223   &0.013 &0.241  &  0.011 & 0.240  & 0.011 \\
pop.total &0.216  & 0.005   & 0.212  & 0.013 &0.263  &  0.011 & 0.317 &  0.012\\
\hline
\end{tabular} 
\caption{Estimate and standard error for each coefficient $\gamma_j$. For the Bayesian models, we report the mean and the standard deviation from 1000 independent draws from the posterior distribution. }
\label{tb:covariates}
\end{table}

\subsection{Random borders for only $\Balpha$\label{randomWalpha}}
Given the smaller number of barriers detected for the $\beta_i$'s compared to the $\alpha_i$'s in Figure 5 of our main paper, we also implemented an alternative model where the adjacency structure $\W^\beta$ for the $\beta_i$'s is considered fixed, and only the adjacency matrix $\W^\alpha$ for the $\alpha_i$'s is allowed to vary.   

The in-sample MSE of 0.0711 for the model with variable $\W^\alpha$ and fixed $\W^\beta$ is worse then the in-sample MSE of 0.0706 for the model with variable $\W^\beta$ and $\W^\alpha$.  However, the variable $\W^\alpha$ and fixed $\W^\beta$ model does have a slightly better out-of-sample MSE of 0.1050 compared to the out-of-sample MSE of 0.1069 for the model with variable $\W^\beta$ and $\W^\alpha$.  

These results provide a further indication that there is stronger signal in the data for detecting discontinuities for the mean level of crime between neighborhoods compared to discontinuities for the trend in crime over time between neighborhoods.  However, we still report the results for the model with variable $\W^\alpha$ and $\W^\beta$ in our main paper, as it provides additional insight and interpretation in Figure 4 of the main paper.  

\subsection{Variability of neighborhood-specific coefficients\label{variability}}
\begin{figure}
\includegraphics[height = 0.8\textheight]{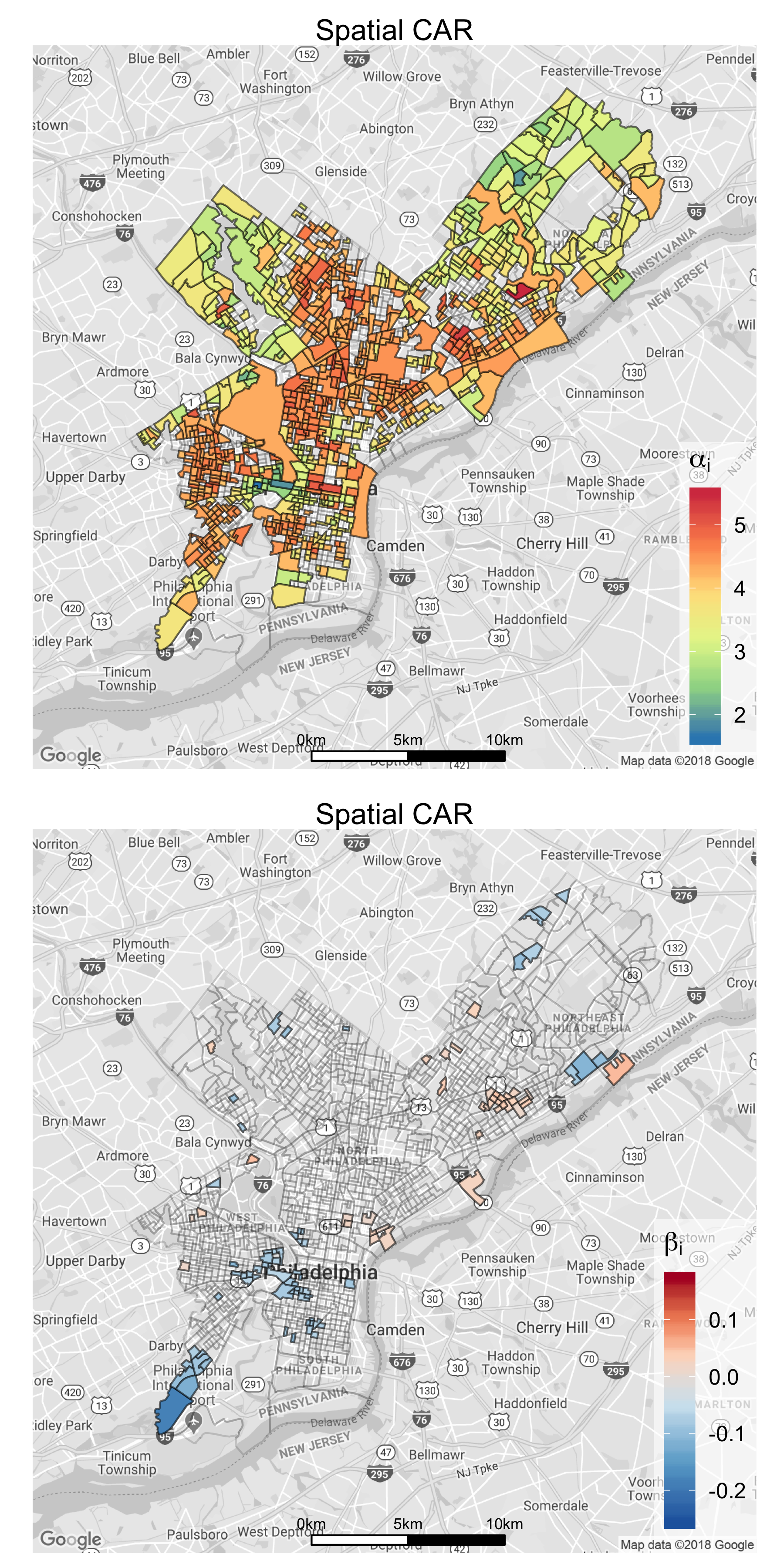} 
\caption{{\bf Top:} The neighborhoods where the $95\%$ credible interval for $\alpha_i$ does not contain the global mean level of crime. {\bf Bottom:} The neighborhoods where the $95\%$ credible interval for $\beta_i$ oes not contain the global time trend in crime. These maps were created with the R package {\tt ggmap} \citep{kahle2013ggmap}.}
\label{fig-sig}
\end{figure}

In Section~5.4 of our paper, we report the neighborhoods with the highest and lowest estimates of the mean level of crime $\alpha_i$ and time trend $\beta_i$.   As a supplement to these results, in Figure~\ref{fig-sig} we map the Philadelphia neighborhoods which are ``significant", in the sense that their $95\%$ credible intervals do not contain the global mean $\bar{\alpha}$ or $\bar{\beta}$ across all neighborhoods in the city. 

In Figure~\ref{fig-sig}, we see many more neighborhoods with significant differences in terms of their $\alpha_i$'s, which is another indication that the variation in the mean level of crime is larger than the variation in the time trend in crime ($\beta_i$'s).  In the plot for $\beta_i$'s, we find a smaller number of neighborhoods with values that are significantly different than the overall mean, but the existence of these neighborhoods confirms the presence of the space-time interaction found in previous studies \citep{law2014bayesian, li2014space}.

\begin{figure}
\includegraphics[height = 0.8\textheight]{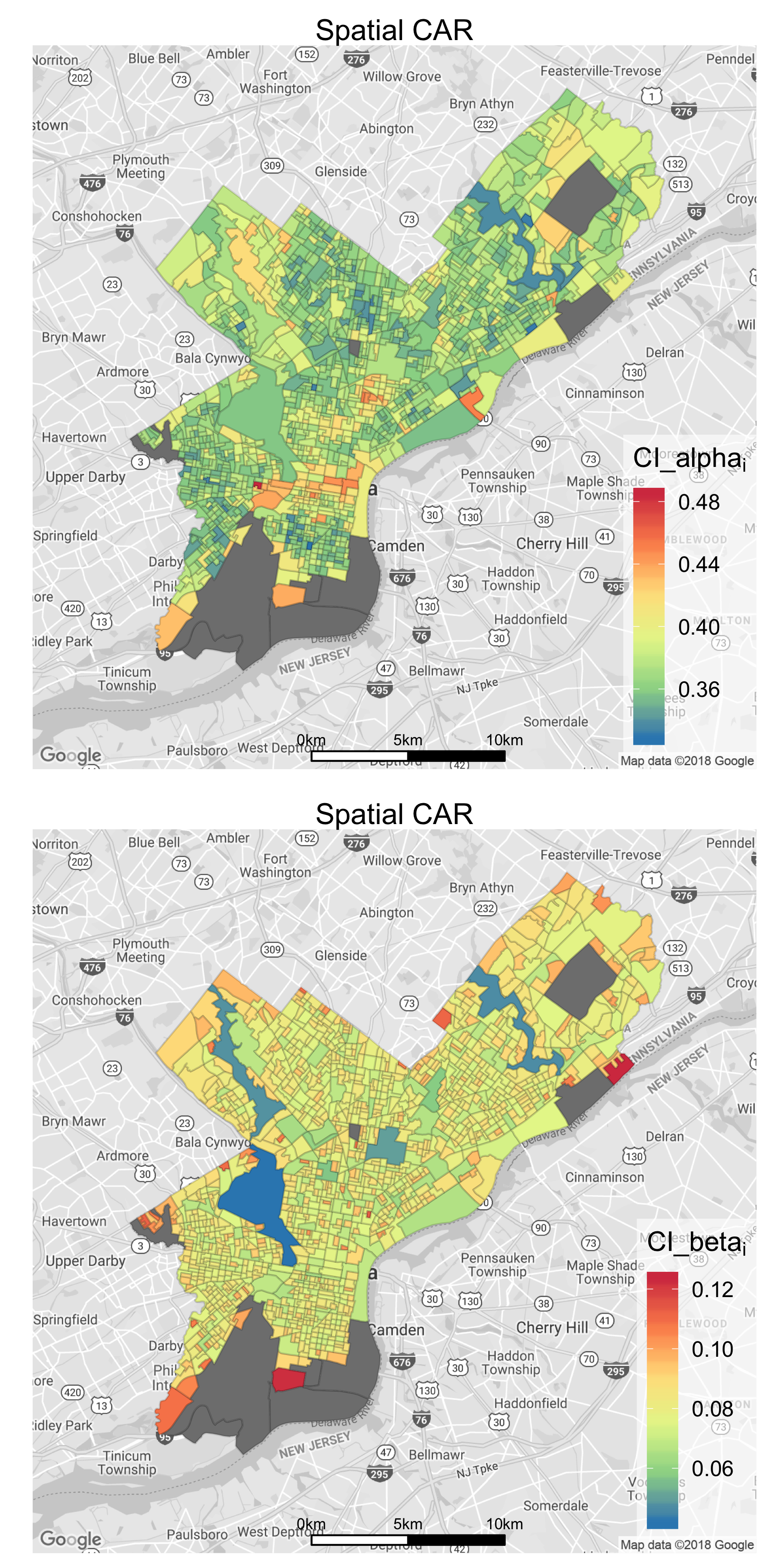} 
\caption{{\bf Top:} The width of the $95\%$ credible intervals for $\alpha_i$. {\bf Bottom:} The width of the $95\%$ credible intervals for $\beta_i$. These maps were created with the R package {\tt ggmap} \citep{kahle2013ggmap}.}
\label{fig-CI}
\end{figure}

In Figure~\ref{fig-CI}, we visualize the width of the $95\%$  credible intervals for each neighborhood-specific $\alpha_i$ and $\beta_i$.    It is interesting to observe that the interval widths are smallest for areal units that border many other units and largest for areal units with very few neighbors.  The neighborhoods with smallest widths are the parks (Fairmount, Wissahickon and Pennipack) which are surrounded by many block groups due to their large surface area. The neighborhoods with largest widths are at the border of the city with only one or two neighboring units.   This phenomenon is more striking for the trends in crime over time ($\beta_i$'s) for which there is a less strong signal in the data compared to the mean level of crime ($\alpha_i$'s) for each neighborhood.

\end{document}